\documentclass[a4paper,fleqn,usenatbib]{mnras}

\usepackage{newtxtext,newtxmath}

\usepackage[T1]{fontenc}
\usepackage{ae,aecompl}

\usepackage{graphicx}	
\usepackage{amsmath}	
\usepackage{amssymb}	
\usepackage{epstopdf}	

\usepackage{pifont}
\newcommand{\xmark}{\ding{55}}%

\title[A three-phase amplification of the cosmic magnetic field]{A three-phase amplification of the cosmic magnetic field in galaxies}

\author[S. Martin-Alvarez et al.]{Sergio Martin-Alvarez,$^{1}$\thanks{E-mail: sergio.martin@physics.ox.ac.uk (SMA)}
Julien Devriendt,$^{1,2}$
Adrianne Slyz,$^{1}$ and Romain Teyssier$^{3}$
\\
$^{1}$Subdepartment of Astrophysics, University of Oxford, Keble Road, Oxford, OX1 3RH, UK\\
$^{2}$Universit\' e de Lyon, Universit\' e Lyon 1, ENS de Lyon, CNRS, Centre de Recherche Astrophysique de Lyon\\
UMR5574, F-69230 Saint-Genis-Laval, France\\
$^{3}$Institute for Computational Science, University of Zurich, Winterthurerstrasse 190, CH-8057 Zurich, Switzerland
}

\date{Accepted 2018 June 14. Received 2018 June 14; in original form 2018 May 9}

\pubyear{2018}

\begin{document}

\definecolor{olive}{cmyk}{0.64,0,0.95,0.40}
\definecolor{orange}{rgb}{1,0.5,0}

\label{firstpage}
\pagerange{\pageref{firstpage}--\pageref{lastpage}}
\maketitle

\begin{abstract}
Arguably the main challenge of galactic magnetism studies is to explain how the interstellar medium of galaxies reaches energetic equipartition despite the extremely weak cosmic primordial magnetic fields that are originally predicted to thread the inter-galactic medium. Previous numerical studies of isolated galaxies suggest that a fast dynamo amplification might suffice to bridge the gap spanning many orders of magnitude in strength between the weak early Universe magnetic fields and the ones observed in high redshift galaxies. To better understand their evolution in the cosmological context of hierarchical galaxy growth, we probe the amplification process undergone by the cosmic magnetic field within a spiral galaxy to unprecedented accuracy by means of a suite of constrained transport magnetohydrodynamical adaptive mesh refinement cosmological zoom simulations with different stellar feedback prescriptions. A galactic turbulent dynamo is found to be naturally excited in this cosmological environment, being responsible for most of the amplification of the magnetic energy. Indeed, we find that the magnetic energy spectra of  
simulated galaxies display telltale inverse cascades. Overall, the amplification process can be divided in three main phases, which are related to different physical mechanisms driving galaxy evolution: an initial collapse phase, an accretion-driven phase, and a feedback-driven phase. While different feedback models affect the magnetic field amplification differently, all tested models prove to be subdominant at early epochs, before the feedback-driven phase is reached. Thus the three-phase evolution paradigm is found to be quite robust vis-a-vis feedback prescriptions.
\end{abstract}






\section{Introduction}
Magnetic fields are one of the principal components of the Universe, with a presence that can be traced over an incredibly wide range of scales, from planets to clusters of galaxies and beyond. While magnetism is not expected to have a considerable impact on the formation process of the large-scale structure of the Universe (LSS), at galactic scales magnetic energy is believed to be in equipartition with the thermal energy in the interstellar medium \citep[ISM,][]{Wolfe92,Bernet08,Beck13b,Mulcahy14}. Therefore, magnetic fields on these scales could impact gas dynamics, and thus emerge as an essential component in the modelling of galaxy formation and evolution. Moreover, observations hint that magnetic fields are critical for several other processes in galaxies. For instance, they are crucial ingredients in the propagation of cosmic rays \citep{Dubois16,Wittor17,Alves-Batista17} and the generation of jets, not to mention key contributors to the processes of star formation \citep{Li09,Hennebelle14,Hull17} and feedback, both stellar \citep{Chen17,Ntormousi17} and from Active Galactic Nuclei \citep{Bambic18}.

State-of-the-art numerical methods are able to model the overall structure of the LSS and the properties of individual galaxies and galaxy clusters with remarkable accuracy \citep[e.g.][]{Vogelsberger14,Hopkins14,AnglesAlcazar14,SantosSantos16,Kaviraj17,Grisdale17}. However, amongst the commonly missing ingredients in simulations of galaxies, magnetic fields stand out for their importance. This can be attributed to their intrinsically three-dimensional divergenceless vectorial nature which significantly increases the complexity of their correct numerical implementation and the daunting dynamical range required to capture their amplification. In an attempt to remedy the situation, various studies have recently explored the issue of the role played by magnetism both in galaxies and the LSS \citep{Dubois10, Gressel13, Beck13a, Hennebelle14, Vazza14, Pakmor14, Vazza15, Gheller16, Marinacci16, Rieder16, Su17, Butsky17}. However, despite these laudable efforts, key questions remain unanswered.

Arguably the most intriguing puzzle about magnetism is the question of its primordial origin, and how its amplitude evolves to reach present-day values. The exact formation time and mechanism of magnetic fields in the Universe is unclear \citep[see][for a review of different magnetogenesis scenarios]{Widrow02}. Theoretical models suggest the formation of weak fields in the very early Universe, typically during the inflationary phase or cosmological phase transitions \citep{Hogan83, Kolb90, Ratra92}. Various astrophysical plasma seeding processes have also been proposed \citep{Biermann50,Schlickeiser12}, suggesting the generation of a magnetic field with strength on the order of $B_0 \lesssim 10^{-19}$ Gauss. These field values are extremely weak, differing by many orders of magnitude from the micro-Gauss fields observed in galaxies and galaxy clusters \citep{Beck13b,Mulcahy14}, and arguably even voids. How is such an enormous gap between primordial and current magnetic fields bridged? Present day averaged cosmic magnetic fields are constrained by observations which provide both lower and upper limits \citep[$\sim 10^{-9} \text{-} 10^{-16}$ G, ][]{Neronov10,Taylor11,Planck16}, although the validity of the lower limit is still debated \citep[e.g.][]{Broderick12}. In any case, the range of values compatible with these observations remains wide, and as such does not provide stringent constraints on the generation mechanism.

For the specific case of galaxies, the gap between primordial and observed field values is the largest. The strength of ordered magnetic fields are typically found to be on the order of $10^{-5} \text{-} 10^{-6}$ G \citep{Beck15,Mulcahy17}. Therefore, extremely effective dynamo mechanisms are required to amplify the primordial fields, especially since equipartition fields are already reported at high redshifts \citep[$z \simeq 2.0$]{Wolfe92,Bernet08}. As a consequence, any dynamo mechanism proposed as a solution needs to be able to amplify magnetic fields during the early stages of galaxy formation and evolution, as well as sustain them to the present day. 
An alternative scenario to the primordial origin of cosmic magnetism, stipulates that astrophysical sources be the progenitors of magnetic fields. These can range from stellar dynamos to active galactic nuclei \citep{Rees87,Vazza17} and supernovae (SNe) \citep{Beck13a,Butsky17}. However, this scenario faces the challenge of how to stretch spatially localised fields to galactic scales and transfer them to the intergalactic medium and LSS. 

Various fast and powerful amplification mechanisms able to strengthen the magnetic field on short time-scales have been invoked to address this problem. Some are the $\alpha-\Omega$ dynamo \citep{Wang09,Dubois10} and the turbulent dynamo \citep{Vazza14,Pakmor14,Rieder16,Pakmor17,Rieder17a,Vazza18}. A magneto-rotational instability-driven dynamo is another possible candidate  \citep{Machida13,Gressel13}, once a field strong enough to excite it has developed \citep{Kitchatinov04}.
Due to the complexity of the problem, numerical studies have become an important tool to probe different dynamos and seeding mechanisms. In fact, most have already been tested in plasma simulations \citep[e.g.][]{Haugen04,Schekochihin04,Haugen04b,Teyssier06,Schober17} and simulations of individual, isolated galaxies \citep{Dubois10,Pakmor13,Rieder16}. Plasma simulations suggest that a turbulent dynamo is efficient enough to amplify the magnetic energy up to equipartition \citep{Schekochihin02,Federrath11b,Federrath14,Schober15}, but such simulations are restricted to scales several orders of magnitude below that of a galaxy. Furthermore, even these simulations are restricted to a limited set of scales, being unable to simultaneously fully resolve both the inertial and subviscous ranges \citep[see][]{Kinney00,Schekochihin02}. Meanwhile, in isolated galaxies simulations, the amplification exhibits moderate growth rates. The main reason is that successfully modelling turbulent dynamo amplification requires extreme spatial resolution to capture a significant fraction of the inertial range and a considerable level of galactic turbulence. The task becomes more arduous when this picture is studied in a cosmological context, where galaxies are born intrinsically small at high redshifts and there is a further added cost for also simulating the cosmological environment. However, rapid magnetic amplification has recently been reported in simulations of both isolated \citep{Rieder16} and cosmological \citep{Rieder17b} dwarf galaxies. 

In this paper, we follow the evolution of the magnetic energy in a Milky Way-like galaxy forming in an explicit cosmological environment with high spatial resolution ($10$pc), sufficient to trigger the turbulent dynamo. As far as we are aware, this is the first time this experiment is performed at such a resolution whilst employing a code that guarantees negligible divergence of the magnetic field at all times. Our work extends the study of magnetic field amplification occurring in isolated spiral galaxies \citep{Rieder16} by accounting for processes such as galaxy mergers and gas accretion. We find that a turbulent dynamo also leads to growth of the magnetic energy, but that the amplification is modulated by environmental effects, thus undergoing different phases where magnetic growth occurs at different rates. Repeating our simulation in turn without stellar feedback and with a stellar feedback prescription different from our fiducial one, we establish that feedback only dominates the amplification process during the later stages of galaxy evolution, when gas accretion and mergers have significantly subsided.

The structure of the manuscript is as follows: the numerical set-up is described in detail in section \ref{s:NumericalMethods}. Results are described in section \ref{s:Results}, where three main epochs of amplification are identified: collapse (section \ref{ss:CollAmplification}), accretion-driven (section \ref{ss:AccAmplification}) and feedback-driven (section \ref{ss:FbAmplification}) amplification. Section \ref{s:Robust} discusses the robustness of this three-phase amplification and reviews implementation-dependent aspects of our findings. Our concluding remarks are given in section \ref{s:Conclusion}. We add two appendices addressing numerical resolution (appendix \ref{ap:resolution}), and presenting the method we use to calculate the Fourier power spectra of a disk galaxy embedded within an explicit  cosmological context (appendix \ref{ap:diskSpec}).

\section{Numerical methods}
\label{s:NumericalMethods}
To run the cosmological zoom-in simulations presented in this paper, we employ a modified version of the publicly available magneto-hydrodynamical (MHD) code  {\sc ramses} \citep{Teyssier02} which accounts for the magnetic field
dilution caused by the expansion of the Universe.  {\sc ramses} couples a tree-based Adaptive Mesh Refinement (AMR) Eulerian treatment of the gas with an N-body treatment for the dark matter and stellar components.  In this section, we briefly introduce the simulation and discuss various numerical aspects relevant for assessing the robustness of our results. 

\subsection{Simulations}
\label{ss:Simulations}

\subsubsection{Initial conditions, resolution \& cooling/heating}
\label{sss:IC}

The initial conditions for all simulations are those of the {\sc nut} suite \citep{Powell11} and consist of a cosmological box of 9 $h^{-1}$ Mpc comoving side, with a 3 $h^{-1}$ Mpc comoving diameter zoom-in spherical region. In this sphere, the code is allowed to refine cells in a quasi-Lagragian way (so that any cell contains roughly the same amount of mass) up to a maximum physical spatial resolution of 10 pc. A $M_\text{vir}\sim 5 \cdot 10^{11} M_\odot$ DM halo is formed in this region by redshift $z=0$. The dark matter and star particle minimal masses are $M_\text{DM} \sim 5 \cdot 10^4 M_\odot$ and $M_{*} \sim 5 \cdot 10^3 M_\odot$ respectively. Initial conditions are set at redshift $z = 500$, with relevant cosmological parameters chosen according to WMAP5 values \citep{Dunkley09}. The simulations also include a UV background to take into account inter-galactic medium photo-ionisation effects \citep{Haardt96} (re-ionisation of the simulated volume is assumed to happen instantaneously at $z=10$), and metallicity dependent gas cooling both at temperatures above $10^4$K \citep{Sutherland93} and below \citep{Rosen95}. The gas is assumed to be monatomic with specific heat ratio $\gamma = 5 / 3$ and follows an ideal equation of state.

\subsubsection{Star formation}
\label{sss:SF}
Star formation is generally modelled in simulations following a Kennicutt-Schmidt law \citep{Schmidt59} with a constant efficiency $\epsilon$. As long as it is employed on large (kpc) scales, the constant efficiency approximation provides a fair description of the observational data. However, at smaller (pc) scales, it is expected that $\epsilon$ will depend on local ISM properties \citep{Padoan11,Hennebelle11,Federrath12}. This motivates a thermo-turbulent star formation model where stars form with a high efficiency ($\epsilon \geq 0.1$) in regions where gravity overcomes the turbulent, magnetic and thermal pressure support. Such a spatially varying star formation efficiency prescription has been introduced in \citep{Kimm17,Trebitsch17,Mitchell18} and is presented in more detail together with an analysis of its impact in Devriendt et al. (in preparation).

All the runs presented in this work spawn star particles in the highest resolution cells \citep{Rasera06} in line with this calculation of the local star formation efficiency. The resulting stellar particles can then generate stellar feedback following one of the two different stellar feedback prescriptions outlined below. A Kroupa initial mass function \citep[][IMF]{Kroupa01} is assumed for the stellar populations. Each supernova returns a mass fraction of gas $\eta_\text{SN} = 0.213$ and newly formed metals $\eta_\text{metals} = 0.075$ back to the ISM which are then advected by the code as passive scalars.

\subsubsection{Mechanical feedback}
\label{sss:Mech}
The first feedback prescription (Mech) is a mechanical supernovae feedback model presented in \citet{Kimm14,Kimm15}. For a given numerical resolution, this model calculates the proper amount of momentum injection by a Sedov-Taylor blast wave which depends on the stage of the explosion. This momentum --- along with energy and mass --- depends on various local parameters (e.g. number of occurring SNe, metallicity and surrounding density of the ISM) and is anisotropically deposited in the cell containing the SN and its neighbours. This approach considerably reduces artificial cooling \citep{Marri03,Slyz05}. See \citet{Kimm14} and \citet{Kimm15} for a detailed explanation of the implementation and \citet{Rosdahl17} for an extended comparison to other popular SN feedback implementations. 

Note that due to its better intrinsic capability to capture the anisotropic expansion of shock waves through local inhomogeneities, this model is expected to better describe shock turbulence. Indeed, it should drive baroclinic vorticity more efficiently and result in higher solenoidal turbulence at small scales. Solenoidal forcing in shocks in turn fosters magnetic amplification \citep{Suoqing16}, without the need for large amounts of energy injection.

\subsubsection{Thermal and radiative feedback}
\label{sss:RdTh}
The second feedback prescription (RdTh) includes both SNe feedback and stellar radiation. Energy injection from SNe alone might not suffice to generate galactic outflows. Stellar radiation has been invoked to supplement the driving power \citep{Murray10}. This model implements such a radiatively boosted stellar feedback, which couples to a standard thermal energy injection by SNe, rendering this latter more efficient. 

More specifically, all star particles in the simulation emit UV photons. Dust in the ISM absorbs these photons, and re-emits the absorbed energy at infrared wavelengths. Infrared photons then transfer momentum to the gas and drive galactic outflows \citep{Murray10,Roskar14}. The efficiency of such a momentum transfer is controlled by the infrared dust opacity $\kappa_\text{IR}$ in our implementation. We set this free parameter to a  value of $\kappa_\text{IR} = 20 \text{ cm}^2 \text{ g}^{-1}$, which mimics an opaque ISM, very efficient at absorbing infrared energy \citep{Semenov03} and generates large outflows. Detail of the implementation can be found in \citet{Roskar14}. SN explosions are modelled as a simple injection of mass, metals and thermal energy into their host cell.

\subsubsection{The simulation set}
\label{sss:set}

The simulation set used in this work consists of three principal runs, B-Mech, B-NoFb, and B-RdTh which are distinguished by their stellar feedback prescriptions, and four supplementary runs performed at lower resolution, B-MR20, B-MR40, B-MR80, and B-MR160. The B-Mech and B-MRX runs all employ the mechanical SN feedback implementation (section \ref{sss:Mech}). Runs labelled MR20, MR40, MR80, and MR160 have a maximal spatial resolution lower than that of their fiducial counterpart B-Mech by factors of $2$, $4$, $8$, and $16$ respectively. The B-NoFb run has no stellar feedback, and the B-RdTh run employs the thermal SN prescription and radiative stellar feedback presented in section \ref{sss:RdTh}. 
In any figure where they are inter-compared, the B-Mech, B-NoFb and B-RdTh simulations are colour-coded blue, red and green respectively.
The main characteristics of the simulations are summarised in Table \ref{table:setups}. Supplementary runs
 performed with a view to estimate the impact of resolution on our results are mostly discussed in appendix \ref{ap:resolution}.
 All simulations have been evolved down to $z=2$.

\begin{table}
\centering
\caption{Simulations used in the manuscript. Co-moving seed magnetic field $B_0$, maximum resolution $\Delta x_{\text{m}}$ and stellar feedback model are indicated for each run.}
\label{table:setups}
\begin{tabular}{l | l l l}
\hline
Simulation & $B_0$ (G)  & $\Delta x_{\text{m}}$ (pc) & Feedback \\
\hline
\textcolor{blue}{B-Mech} & $3 \cdot 10^{-20}$ & $10$ & Mech (sect \ref{sss:Mech}) \\
\textcolor{red}{B-NoFb} & $3 \cdot 10^{-20}$ & $10$ & \xmark \\
\textcolor{green}{B-RdTh} & $3 \cdot 10^{-20}$ & $10$ & RdTh (sect \ref{sss:RdTh})\\
\hline
B-MR20 & $3 \cdot 10^{-20}$ & $20$ & Mech \\
B-MR40 & $3 \cdot 10^{-20}$ & $40$ & Mech \\
B-MR80 & $3 \cdot 10^{-20}$ & $80$ & Mech \\
B-MR160 & $3 \cdot 10^{-20}$ & $160$ & Mech \\
\hline
\end{tabular}
\end{table}

\subsection{Magnetic field specifics}
\label{ss:MagField}

\subsubsection{Constrained transport}
\label{sss:CT}

The MHD solver of  {\sc ramses} is based on the Constrained Transport \citep[CT][]{Evans88} scheme \citep{Teyssier06,Fromang06}. It solves the induction equation
\begin{equation}
\frac{\partial \vec{B}}{\partial t} = \vec{\nabla} \times \left( \vec{v} \times \vec{B} \right)  +  \eta \vec{\nabla}^2 \vec{B},
\label{eq:Induction}
\end{equation}
(with $\vec{v}$, gas velocity; $\eta$ magnetic diffusivity) to evolve the magnetic field $\vec{B}$ in time, guaranteeing that the solenoidal constraint
\begin{equation}
\vec{\nabla} \cdot \vec{B} = 0,
\label{eq:Solenoidal}
\end{equation}
holds to numerical precision. 
As magnetic diffusivity is negligible on galactic scales, we set $\eta = 0$, therefore all magnetic diffusivity is exclusively due to numerical effects. 
The divergence free aspect of the scheme is especially relevant at shocks and in regions where turbulent motions dominate, as magnetic field tangling may generate large artificial divergences \citep{Balsara04}. Non-constrained methods or divergence cleaning algorithms can typically display a divergence per cell $(\vec{\nabla} \cdot \vec{B})\, \text{dx}_\text{cell}$ on the order of the strength of the local field $|\vec{B}|$, with $\text{dx}_\text{cell}$ the cell size in such regions \citep{Wang09,Kotarba11,Pakmor13}. The main problem is that these highly turbulent regions are also the places where most of the magnetic field amplification is expected to occur if a small-scale, fast dynamo is the culprit. For non-constrained methods, the magnetic mono-polar contribution is preserved once generated, casting serious doubts on their ability to reliably capture any physical amplification mechanism. Divergence cleaning codes base their solenoidal constraint fulfilment on a rapid diffusion of the spurious numerical contribution to the divergence \citep[see][]{Powell99, Dedner02}. However, by the time such a diffusion has brought the divergence to an acceptably low level, the mono-polar contribution has already likely had an impact on both the amplification and the structure of the field (see \citet{Mocz16} where divergence cleaning and CT schemes are compared for an isolated spiral galaxy simulation). We refer the reader to \citet{Hopkins16} for a comparison of how different numerical methods perform in a variety of standard MHD test cases.

\begin{figure}%
\includegraphics[width=\columnwidth]{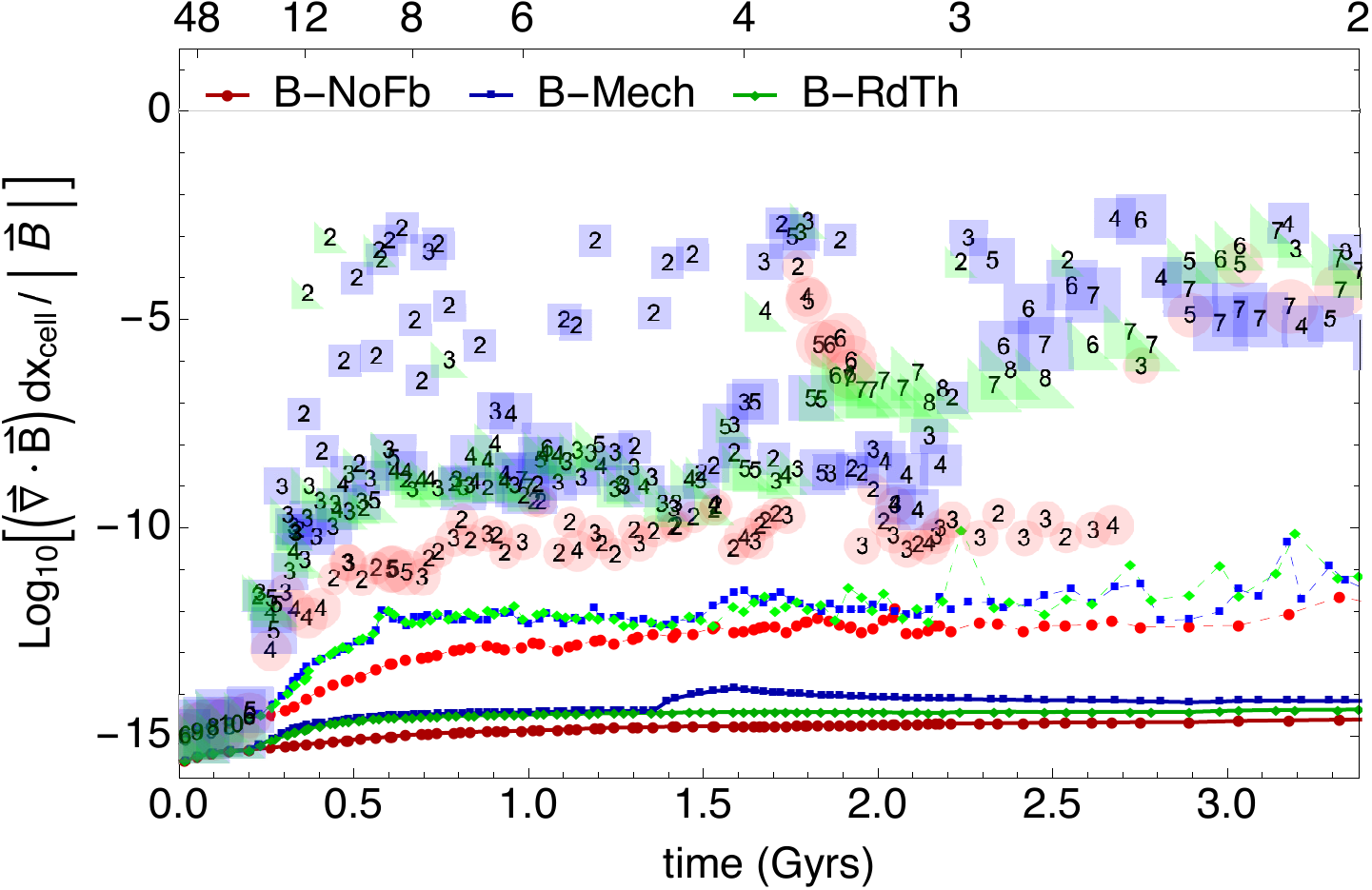}
\caption{Magnetic field divergence to total magnetic field strength ratio $(\vec{\nabla} \cdot \vec{B})\, \text{dx}_\text{cell} / |\vec{B}|$ per cell with length $\text{dx}_\text{cell}$, as a function of redshift. Median (solid lines) and $3 \sigma$ (dashed lines) values of the AMR cell distribution are shown for the entire simulation volume. Squares (B-Mech), circles (B-NoFb), and triangles (B-RdTh), indicate maximal values reached in any individual cell across the whole simulation volume, and are labelled with the level of resolution of that cell. Number 1 corresponds to the highest resolution level, and larger numbers indicate ever lower resolution levels. Note that even those individual maxima always lie below two-three orders of magnitude below the field strength (gray line).}%
\label{CleanDivB}%
\end{figure}

To quantify the accuracy of our CT scheme in the context of cosmological AMR simulations of galaxy formation, we show in Figure \ref{CleanDivB} the values of the divergence to magnetic field ratio $(\vec{\nabla} \cdot \vec{B})\,\text{dx}_\text{cell} / |\vec{B}|$ as a function of redshift. All three main runs presented in sub-section \ref{ss:Simulations} are shown. They display insignificant median values (dark solid lines) of this ratio, at a level consistent with numerical precision. Interquartile ranges are narrow and always found within $\pm 1$ dex above and below the median. Dashed lines indicate the $3 \sigma$ width of  the entire distribution of AMR cells: they are always below a ratio of $10^{-10}$. Finally, large circles, triangles and squares represent the maximal values ever reached by a single cell. These are labelled with the level of refinement of the cell, with 1 corresponding to maximal resolution and higher numbers indicating ever coarser refinement levels. It can be seen that at all times, and for all cells, the divergence to magnetic field strength ratio occurring remains below percent levels.

\subsubsection{Resolution requirements}
\label{sss:Dyn}
\begin{figure}%
\includegraphics[width=\columnwidth]{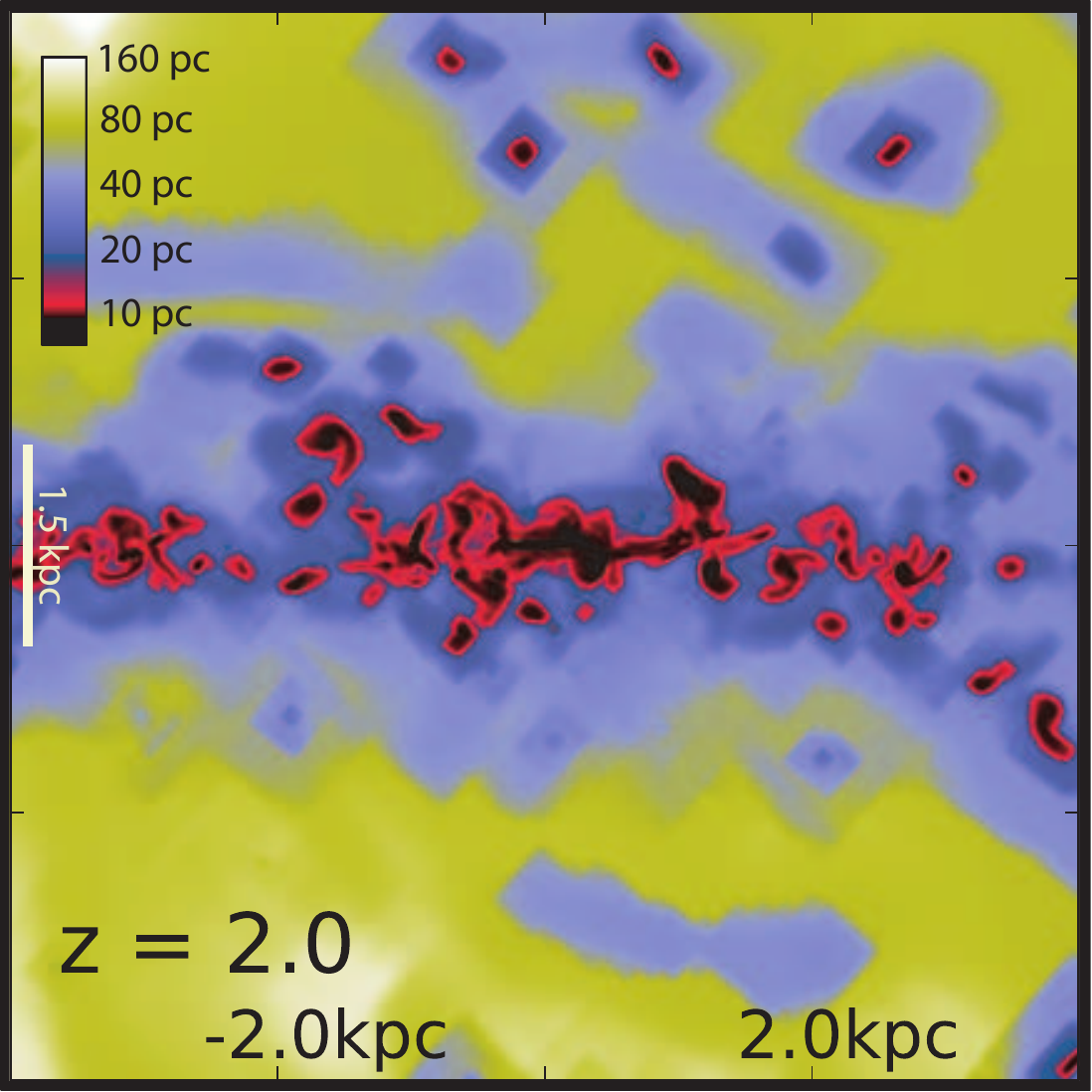}
\caption{Edge-on projection view of a $\left(8 \text{ kpc}\right)^3$ region centred on the B-Mech galaxy at $z = 2$. Colours display density-weighted cell sizes. Neglecting the maximal level of resolution (10 pc, in black) only found in the densest regions, the galactic disk is resolved with approximately $\sim 70$ resolution elements (20 pc, dark blue), well above the necessary threshold advocated by \citet{Federrath11} to achieve 
significant dynamo action.}%
\label{GalRes}%
\end{figure}

It has also been argued by \citet{Federrath11} that self-gravitating MHD simulations require a minimum of 30 resolution elements per Jeans length to achieve significant dynamo action. Due to the multiphase nature of the ISM and the different impact of each phase on the magnetic energy evolution \citep{Evirgen17}, attaining this resolution across as much of the galaxy as possible is of key importance. For spiral galaxies, such a feat becomes rapidly challenging given how thin their gas disk become. Figure \ref{GalRes} displays the typical resolution reached in the disk of our simulated galaxies. The half-thickness $h_s$ estimated from fitting an exponential density profile $\rho (z) \propto e^{- | z | / h_s}$ to the galaxy gas disk in the range of redshift probed by the simulations ($2 < z < 4$), is typically $h_s \sim 0.6$ kpc for the runs with feedback, and $h_s \sim 0.4$ kpc for the run without feedback. The vertical dimension of the galaxy disk thus contains approximately $70$ resolution elements already for the second highest level of resolution. For a thickness of the disk comparable to its Jeans length, this quantity is well above the threshold suggested by \citet{Federrath11}. In Appendix \ref{ap:resolution}, we also discuss how the magnetic Reynolds number requirements to capture a turbulent dynamo are met in our simulations. As far as we are aware, this work is the first to combine this kind of resolution with an accurate CT scheme for MHD in cosmological simulations of disk galaxies.

\subsubsection{Initial conditions}
\label{sss:ICB}
Finally, due to the absence of battery terms in equation \ref{eq:Induction}, magnetic fields have to be initially seeded in the simulations. 
Given the magnetic seed strengths employed, the spatial distribution of the initial magnetic seed has no dynamical effects in collapsing 
proto-galaxies. \citet{Marinacci15} have also shown that the orientation of the initial seed has no impact on the field structure and its average properties within haloes. In virialized regions such as galaxy clusters and galaxies, all memory of the initial seed is also expected to be lost once enough turbulence and amplification have occurred \citep{Dubois10}. As a consequence, the use of a uniform initial field, both in strength 
and orientation has become a standard amongst astrophysical simulations, and we follow suit by using a uniform field oriented along the z-axis of the box. All runs are seeded with an identical, weak magnetic field, assuming a typical Biermann battery seed strength \citep{Biermann50}. This corresponds to a co-moving magnetic field $B_{\text{co}} = 3 \cdot 10^{-20}$ G, which is the natural quantity evolved in the simulations.
Indeed, we introduce a super-comoving magnetic field in  {\sc ramses}, closely following the implementation suggested by \citet{Martel98}.
Our value of  $B_{\text{co}} = 3 \cdot 10^{-20}$ G corresponds to a physical field strength of $B_{\text{ph}} \sim 10^{-17} \text{-} 10^{-18}$ G at $z \sim 40$, in keeping with classical Biermann battery estimates quoted by \citet{Pudritz89,Widrow02}. The selection of such low magnetic seeds ensure that our simulations remain in the kinematic regime of dynamo amplification even after the galaxy undergoes a significant initial compression. Therefore, we can focus on measuring the amount of magnetic turbulent dynamo amplification and whether it leads to saturation. 

We stress, however, that implementing a different seeding (structure of the field) even with our weak primordial magnetic field strength might lead to a different amplification level during the early collapse phase. For instance, \citet{Rieder16} measure such an effect in their simulations of 
isolated galaxies. We defer a detailed investigation of this issue to future work.

\subsection{Halo and galaxy identification}
\label{ss:HaloFinder}
Any study of galaxies in cosmological simulations requires a clear discrimination between these objects and their environment. In our simulations, this is achieved by running the HaloMaker software described in \citet{Tweed09}, on the dark matter particle distribution. The resulting structures, obtained via a thresholding of the dark matter density field, are identified as dark matter halos. Note that sub-halos are 
also identified in this way using the AdaptaHOP algorithm \citep{Aubert04}. Within these dark matter halos, the most massive bound baryonic 
objects are then detected and identified as galaxies.

Various parts of the analysis presented in this paper require an accurate identification of galaxy centres (e.g. 
measuring rotational velocities or generating density/velocity/energy profiles). The centre of our bound baryonic structures are 
calculated according to the shrinking sphere method proposed by \citet{Power03} applied to gas and stars (when these latter are present).
Finally,  unless otherwise stated, all physical quantities are measured within a sphere centred on each galaxy and extending out to $0.2\,r_{\text{halo}}$, where  $r_{\text{halo}}$ stands for the virial radius of the halo. 

\section{Results}
\label{s:Results}
In this section, we evaluate magnetic energy growth in simulated galaxies. As a preamble, we note that, by the end of the simulated period, i.e. $z = 2$, magnetic fields in observed galaxies have already reached equipartition with other forms of energy \citep{Bernet08}. Thus 
we expect that the most significant part of magnetic energy amplification occurs sometime between the seeding of the primordial fields and this redshift.

\begin{figure*}%
\includegraphics[width=2.\columnwidth]{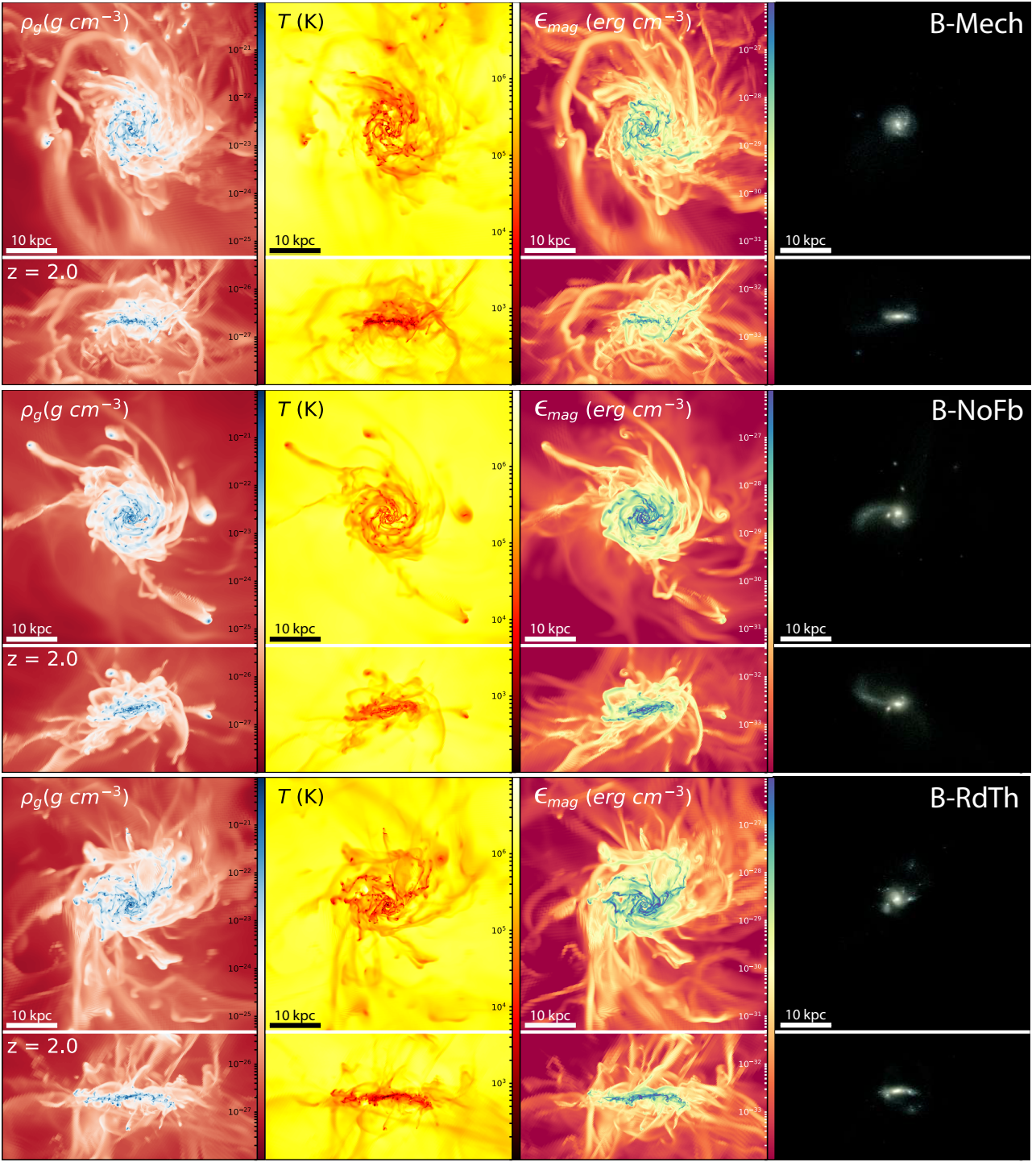}
\caption{Projections of various variables centred on the main galaxy at z = 2, for the different simulations: B-Mech (top), B-NoFb (middle) and B-RdTh (bottom). Panels show face-on and edge-on views of each galaxy, in $\left(\text{50 kpc}\right)^3$ boxes. From left to right: gas density $\rho_g$ (g cm$^{-3}$), temperature $T$ (K), magnetic energy density $\epsilon_\text{mag}$ (erg cm$^{-3}$), and mock optical images. Mock graphs are obtained by retrieving all the stellar particles inside the displayed volume and computing their emission in the SDSS [u',g',r'] filters using \citet{Bruzual03} Single Stellar Population (SSP) models. For this calculation we use the code {\sc sunset}, which is simplified version of the {\sc stardust} algorithm \citep{Devriendt99}. Intensive quantities are mass weighted.}%
\label{Galaxies}%
\end{figure*}

In contrast to simulations of isolated galaxies, galaxies formed within an explicit cosmological context evolve following a hierarchical formation scenario. More specifically, Fig. \ref{Galaxies} presents the gas density, temperature, magnetic energy density and optical mock images for our galaxies at $z = 2$.  In reaching this epoch, galaxy properties have been sculpted by various environmental effects such as cold filamentary and hot diffuse modes of gas accretion and mergers. As galaxies grow, the evolution of their magnetic fields is influenced by such changing environment. We identify three different phases in the evolution of the galaxy magnetic field, from the beginning of the simulations to redshift $z = 2$. These three phases  are well correlated with different physical processes governing the evolution of the galaxy. They are: the initial collapse of the proto-galaxy, an accretion phase during which the main 
factor affecting galaxy properties is the fuelling of its growth by strong cold flow accretion, and a feedback-dominated phase where stellar
feedback finally overcomes the dwindling accretion. During each phase, we assess the effects of stellar feedback and cosmological environment processes.

All quantities in this paper are expressed in physical units (by opposition to co-moving) unless otherwise indicated.

\subsection{Formation of the galaxy: the collapse phase}
\label{ss:CollAmplification}
\begin{figure}%
\includegraphics[width=\columnwidth]{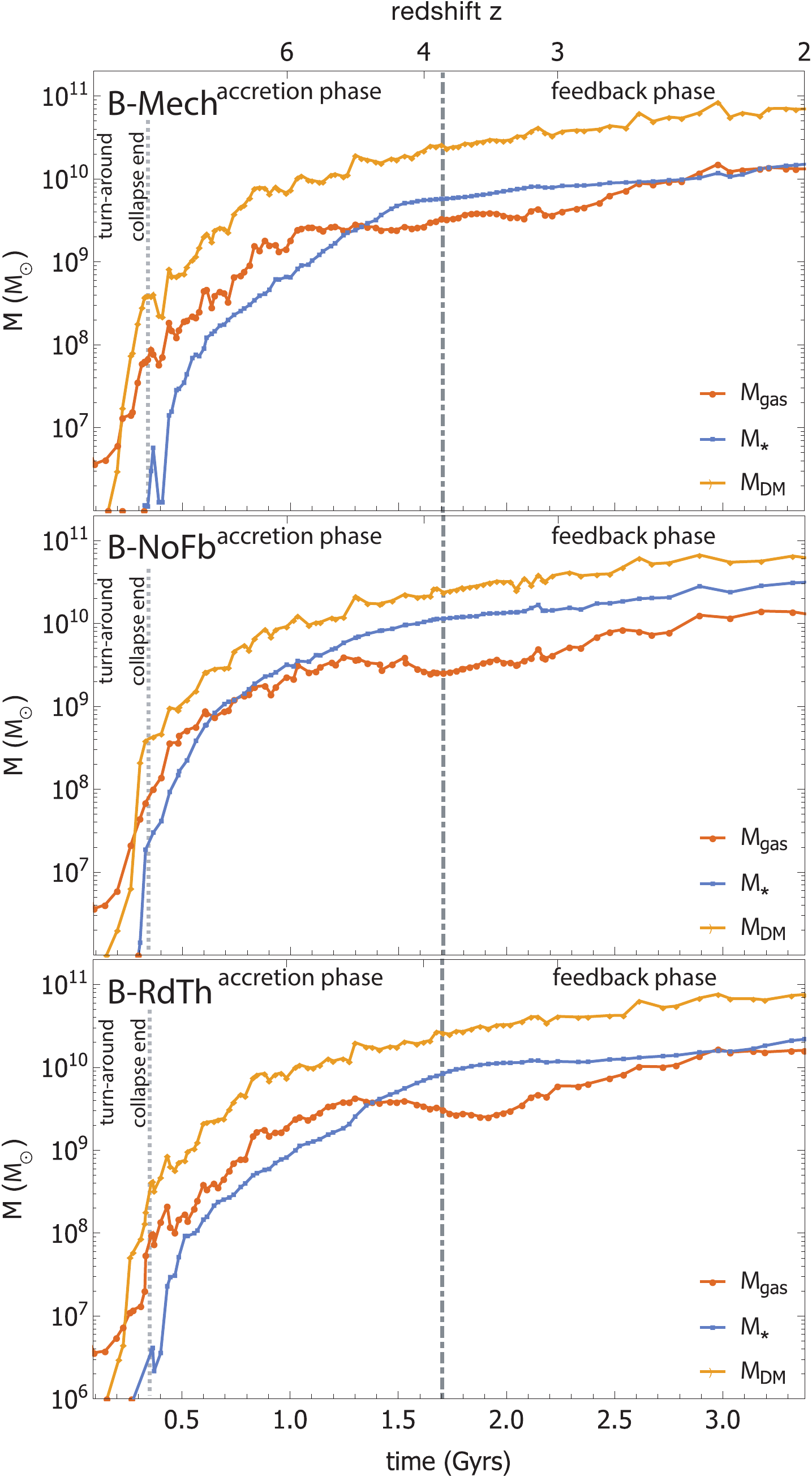}
\caption{Gas (red), dark matter (yellow), and stellar (blue) masses within the galactic region for the B-Mech (top), B-NoFb (middle) and B-RdTh (bottom) runs. Vertical lines identify the end of the collapse (dotted) and the accretion (dot-dashed) phases.}%
\label{MassEvolution}%
\end{figure}

The growth of a galaxy in a cosmological cold dark matter simulation follows a hierarchical collapse scenario \citep[e.g.][]{White91}. Initial perturbations expand until they reach the turn-around point, when their magnetic energy reaches a minimum. At this stage, they decouple from the expansion and their collapse begins. The deepening gravitational potential builds up the galaxy by accumulating baryonic mass until stars 
form. In our simulations, we identify the end of the collapse phase ($t_\text{coll}$ or $z_\text{coll}$) with the moment when a substantial amount of star particles (i.e. at least 100) has been created, to give the galaxy enough time to build a multi-phase ISM. Figure \ref{MassEvolution} displays the mass growth experienced by the different mass components within a sphere of radius 0.2 $r_{\text{halo}} (z)$ centred on the main galaxy in the simulations. Note that $r_{\text{halo}} (z)$ represents the physical virial radius of the dark matter halo, which increases as redshift decreases and that the multiplicative constant in 
front of it simply defines a spherical region containing the galaxy and its close surroundings. The value of this constant is based on the 
estimated limiting size of the galaxy disk, $r_\text{gal}$, if it shared a similar amount of specific angular momentum to the dark matter enclosed within the virial radius of its host halo. Indeed, this yields $r_\text{gal} \sim  4 \times \lambda \;  r_\text{halo}$  \cite[eq.14]{Mo98} given the measured spin parameter of our halo, as defined in \citet{Bullock01} remains close to $\lambda \sim 0.05$ at all times \cite{Kimm11}. The central region thus defined will be referred to as the {\em galactic region} in the remainder of the paper. Finally, at very early epochs, i.e. between the turn-around time when the density perturbation decouples from the expansion of the Universe and the formation of the first virialized dark matter halo we are able to detect given our resolution, we use a fixed galactic region physical size 
$r_\text{gal} = 0.7$~kpc~$\simeq 0.2 \; r_\text{halo}(z_\text{coll})$, centred on the co-moving position of the collapsed proto-galaxy at 
the end of collapse. 
We have checked that reasonable alterations to this somewhat arbitrary choices for size and centre of the galactic region do not significantly 
impact our results, as all density fields tend to become very rapidly uniform at these early epochs.

Changes in galaxy mass due to accretion, mergers and outflows lead to ambiguities when comparing absolute energies within the galactic region. Similarly, energy densities are sensitive to the shape and volume of the galaxy, as magnetic lines are frozen in the fluid in ideal MHD. This is especially relevant for the study of disk galaxies. To circumvent this issue, we therefore resort to using specific energies
\begin{equation}
	\varepsilon_{x} = \frac{E_\text{x}}{M_\text{gas}},
	\label{eq:sE}
\end{equation}
obtained by taking the ratio between total energies $E_x$ ({\em not} energy densities, $e_x$, although the two quantities are, of course, related by a simple volume factor) and gas mass $M_\text{gas}$ within the given region. This allows us to disentangle the amplification of the magnetic energy caused by the growth of the galaxy from that of an intrinsic dynamo effect.

\begin{figure}%
\includegraphics[width=\columnwidth]{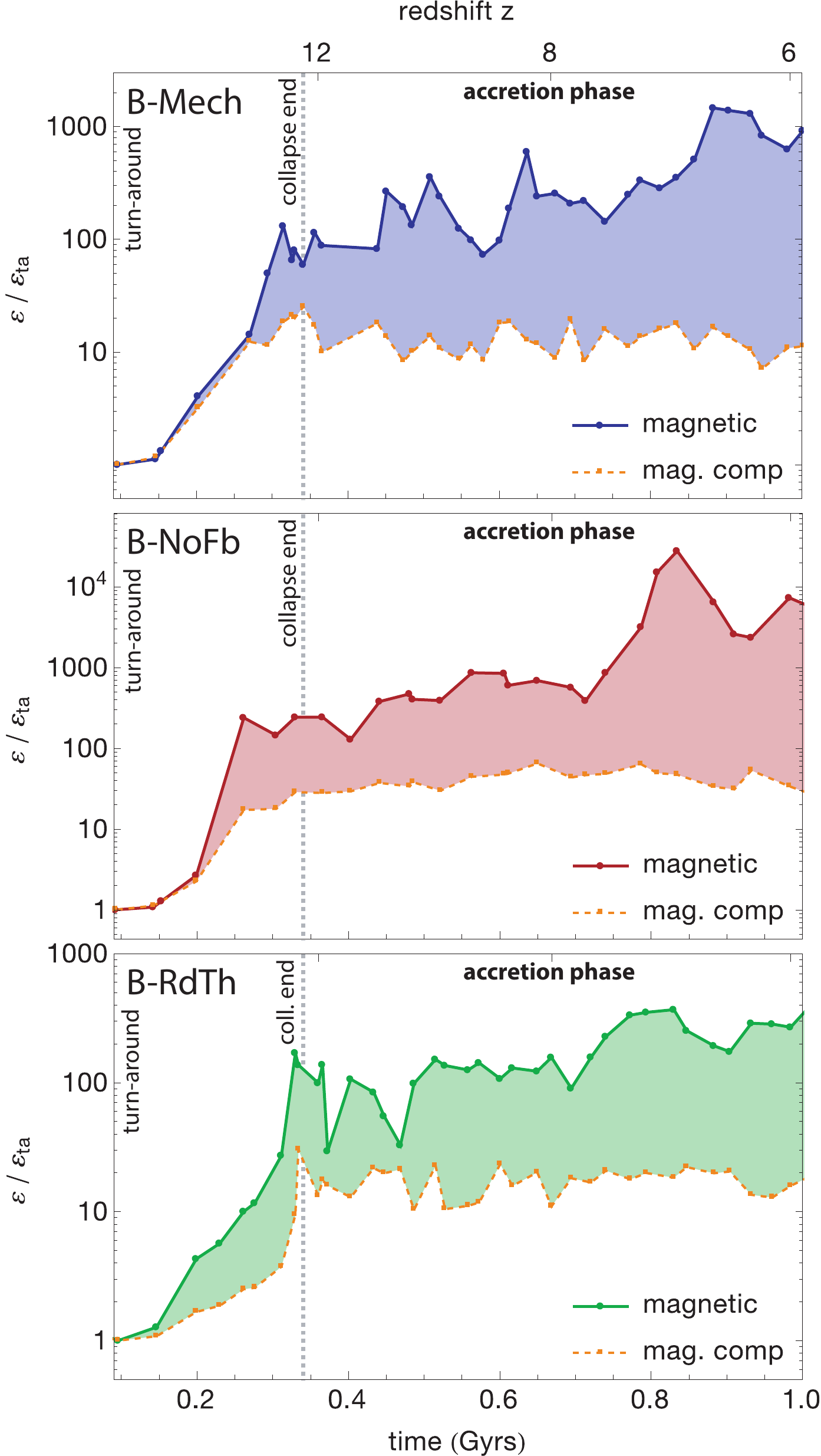}
\caption{Specific magnetic energy growth during the collapse phase and the beginning of the accretion phase within the galactic region (inner $0.2r_{\text{halo}}$ of the dark matter halo). The three main runs B-Mech (blue), B-NoFb (red), and B-RdTh (green) are displayed. Dashed orange lines represent the expected growth of the specific magnetic energy if the collapse were to proceed adiabatically and isotropically. All runs are normalized to their value at the epoch of turn around $t_{\text{ta}}$ (see text for detail). The vertical dotted line marks the end of the collapse phase.}%
\label{EarlyEnergy}%
\end{figure}

During the initial phase of formation, the growth of the magnetic energy density in an idealized, isolated halo closely follows 
an isotropic adiabatic approximation as the gas collapses, so that $e_\text{mag} \propto \rho_g^{4/3}$ where $\rho_g$ is the gas density \citep[e.g.][]{Wang09,  Dubois10, Rieder16}. However, the cosmological formation of a galaxy involves a more complex accretion/collapse history which fosters higher turbulence levels than the isolated case \citep{Klessen10, Elmegreen10}. The relevance of accretion-driven turbulence increases when the growth of the gravitational potential is taken into account \citep{Elmegreen10}. Fig.~\ref{EarlyEnergy} displays the evolution of the specific magnetic energy $\varepsilon_\text{mag}$ within the galactic region during the collapse phase and the beginning of the accretion phase. Note that specific magnetic energies are normalized to their respective (but only different by a few percent) values, $\varepsilon_\text{mag,ta}$, within the galactic region at the time of turn-around for each of the runs. 
As expected,  the collapse times of halos and thus the magnetic amplification levels reached during collapse, are quite robust vis-\`a-vis different stellar feedback prescriptions, at the resolution reached in our simulations.

\begin{figure}%
\includegraphics[width=\columnwidth]{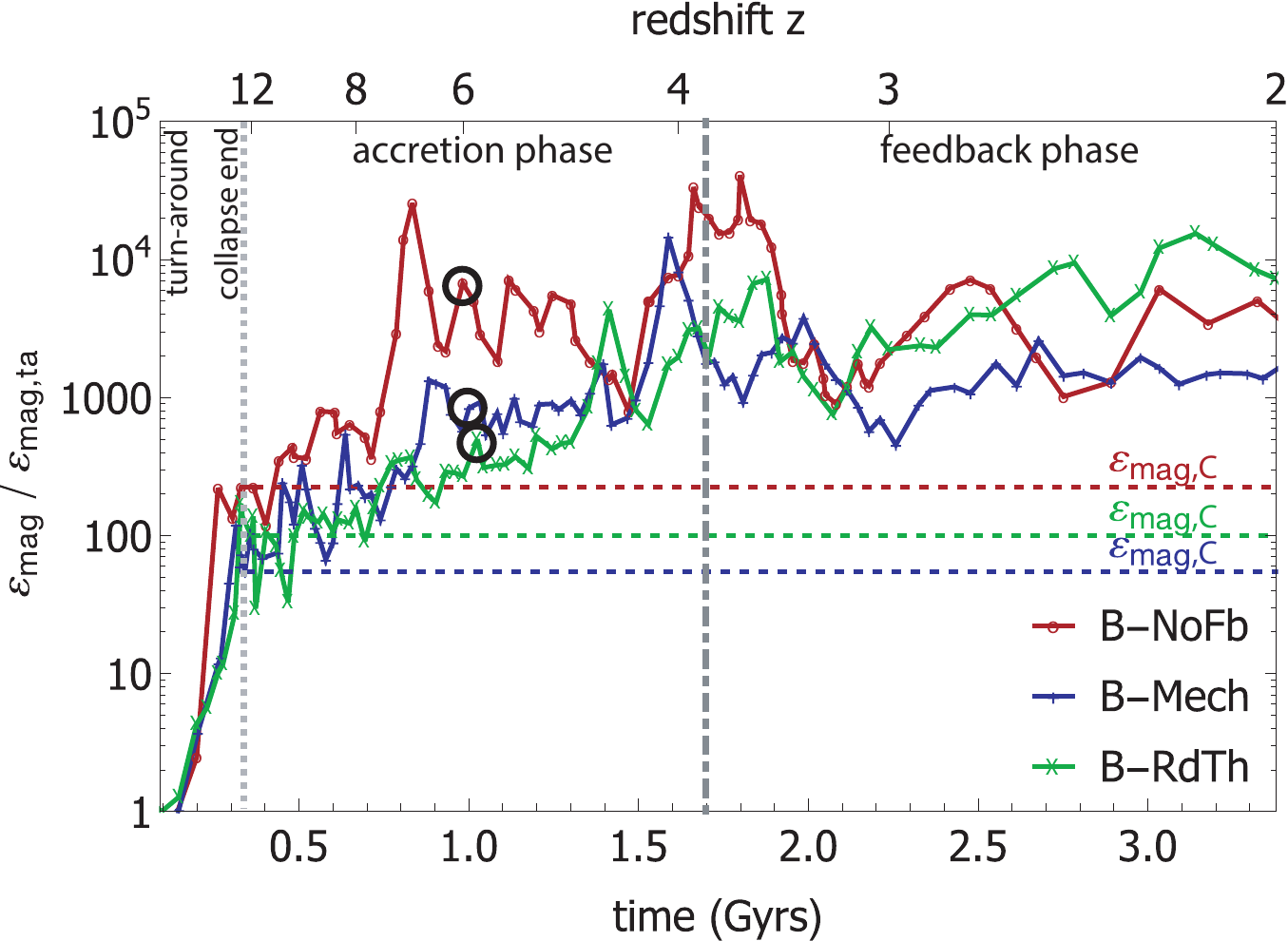}
\caption{Specific magnetic energies growth (solid lines) in the galactic regions for the B-Mech (blue), B-NoFb (red), and B-RdTh (green) runs. Dashed lines show the specific magnetic energies at the end of the collapse phase $\varepsilon_{\text{mag},C}$ for each run. Vertical lines correspond to the beginning of the accretion (dotted) and feedback (dot-dashed) phases. All energies are normalized to their turn-around value.}%
\label{SpecificEmag}%
\end{figure}

More quantitatively, the estimate of the collapse time 
\begin{equation}
	t_{\text{coll}} = \frac{2}{3 H_0}\frac{1}{(1 + z_{\text{coll}})^{3/2}}.
	\label{eq:tCollapse}
\end{equation}
where $z_\text{coll}$ is defined as the redshift at which a spherical top hat density perturbation reaches the linearly extrapolated critical over-density contrast in an Einstein-de Sitter universe \citep[e.g.][]{Peebles94} can be compared with the onset of star formation in simulated galaxies. 
We find the value of the collapse time defined in this manner, $t_{\text{coll}} \sim 0.34$ Gyrs ($z_\text{coll} \sim 13$) (and by extension the turn-around time $t_\text{ta} \simeq 0.5 \; t_\text{coll}$; $z_\text{ta} \sim 22$) to be in fair agreement with the measured 
galaxy formation time (compare the vertical dotted line on Fig.~\ref{EarlyEnergy} to the earliest peak of the dashed curves). Specific magnetic energies measured in the galactic region at exactly $t_\text{coll}$ are used as reference in the rest of the paper and denoted $\varepsilon_{\text{mag},C}$. They provide an upper limit of the magnetic amplification expected from the collapse phase because 
the expansion of the Universe ensures that magnetisation of any gas subsequently fed to the halo will be lower. They are indicated by horizontal lines on Fig.~\ref{SpecificEmag} and spread out by a factor 4 between runs (with values of 55 $\varepsilon_{\text{mag,ta}}$ for B-Mech, 220 for B-NoFb and B-RdTh in between), 
hinting that stellar feedback might play a role in reducing magnetic field amplification in the later stages of the collapse. 
Note that if we relax the collapse time criterion to some degree, allowing the halo in the different runs to collapse at slightly 
different times (which is bound to happen given the chaotic nature of the system) and identify $\varepsilon_{\text{mag},C}$ with the 
peak in specific magnetic energy for each run (solid curves), the gap between runs is still present, even though it is reduced to a factor 1.7 
(130 $\varepsilon_{\text{mag,ta}}$ for B-Mech, 220 for B-NoFb and B-RdTh in between, see Fig.~\ref{EarlyEnergy}). 

From Fig.~\ref{EarlyEnergy}, it is clear that in the simulations, specific magnetic energies $\varepsilon_\text{mag}$ grow by a factor $\sim 100-200$ between the time of turn around (beginning of the time axis) and the end of the collapse phase (vertical dotted lines). These values are about an order of magnitude above those derived in the analytical study of \citet{Lesch95}, who use the spherical top hat model to estimate that the magnetic field strength is amplified by $f_c \sim \left[ r_\text{ta} / (0.2 r_\text{halo}) \right]^2 \sim 100$ from turn-around
until disk formation. Indeed, this corresponds to a specific magnetic energy increase of $ f_c^2 \left[ (0.2 r_\text{halo})/ r_\text{ta} \right]^3 \sim 10$ in the galactic region, --- the gas mass within the sphere of radius $r_\text{ta}$ is assumed to be conserved exactly in the analytic estimate ---, which matches quite well  our own  estimate of the evolution of the specific magnetic energy expected from pure isotropic adiabatic collapse (orange dashed curve in Fig.\ref{EarlyEnergy}). This latter estimate is obtained assuming that each individual cell in the galactic region has undergone such a collapse from the time of turn-around, i.e. that its magnetic energy is given by 
\begin{equation}
	E^\text{cell}_\text{mag} (t) = \frac{\text{dx}_\text{cell}^3}{\frac{4}{3} \pi r_\text{gal}^3(t_\text{ta})} \times \langle E_\text{mag} (t_\text{ta})  \rangle \left( \frac{\rho^\text{cell}_g(t)}{ \langle \rho_g(t_\text{ta}) \rangle } \right)^{4/3}  ,
\end{equation}	
where $\langle \rangle$ stands for averages over the whole galactic region at turn around. One then obtains the adiabatic isotropic estimate for the specific magnetic energy of the galactic region at time $t$ by adding the contributions of all the cells enclosed within the region at that time and dividing it by the total amount of gas present in the region. 
 
 We emphasize that such an estimate not only assumes that the collapse is adiabatic and isotropic, but also that no decrease of the field occurs due to cosmological expansion. In reality, the strength of the magnetic field should be lower post-collapse because (a) the expanding Universe reduces the magnetisation of material accreted later ($B \propto a^{-2}$) and, (b) even though the initial collapse might be close to isotropic, material will subsequently be fed to the galaxy through filamentary accretion and mergers. Filamentary accretion is expected to bring in material with lower magnetisation due to the simple spatial trajectories followed by the infalling gas, given the low amount of cold streams turbulence expected at the mass range of our halo \citep{Mandelker17}. The impact of mergers on magnetic amplification is more complex and its treatment is deferred until section \ref{ss:AccAmplification}. In any case, shortly before the initial collapse phase ends ($t \gtrsim 0.25$ Gyr), this crude estimate of isotropic and adiabatic amplification starts to systematically underestimate the measured specific 
 magnetic energies by a significant amount (solid curves in Fig.~\ref{EarlyEnergy}). In practice, it is found to oscillate  
at a level between $\varepsilon_\text{mag} \sim 10 \text{-} 20\,\varepsilon_\text{mag,ta}$ throughout the simulations (i.e. until $z=2$), and is thus more representative of a configuration where the galaxy accretes material with constant specific magnetic energy throughout its life time, i.e. without any kind of amplification or decay of the magnetic field beyond the initial compression.

In summary, our simulations clearly show (Fig.~\ref{EarlyEnergy}) that by the end of the collapse phase, the level of turbulence induced in the proto-galaxy has already amplified the specific magnetic energy within the galactic region by approximately an order of magnitude above the isotropic adiabatic collapse prediction. Such a significant deviation has also been reported for larger virialized objects such as galaxy clusters \citep[e.g.][]{Dolag05, Dubois08, Marinacci15}. Smaller structures, however, generally still seem to follow the $e_\text{mag} \propto \rho_g^{4/3}$ scaling, even  in the most recent work of \citet{Pakmor17}, which can probably be attributed to their lower spatial and mass resolution (based on the resolution 
study provided in Appendix~\ref{ap:resolution}). Although not easily gathered by looking at Fig.~\ref{EarlyEnergy}, towards the end and immediately after the collapse phase ($0.25\lesssim t \lesssim 0.5$~Gyr), the deviation from the isotropic case is mainly driven by turbulent accretion. Indeed, Fig.~\ref{SpecificEmag}, where we plot the three runs together on the same panel, presents evidence that the more powerful the feedback prescription (the B-RdTh run being the most powerful), the lower the difference with the isotropic collapse estimate. 

\begin{figure}%
\includegraphics[width=\columnwidth]{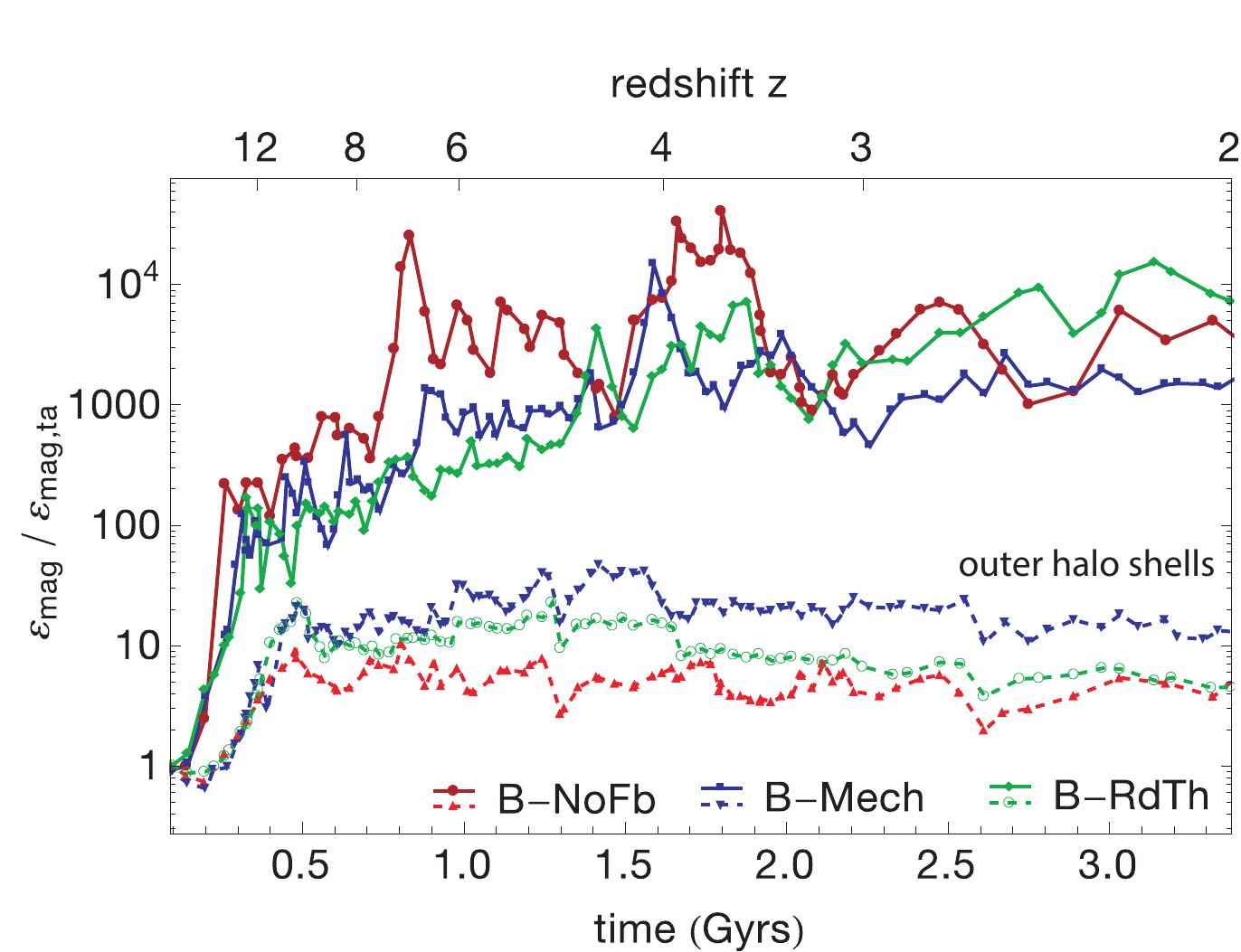}
\caption{Time evolution of specific magnetic energies in a thin shell around the galaxy halo ($r_\text{halo} < r < 1.05\, r_\text{halo}$, dashed lines) for the three main runs compared to the same quantities within the galactic region (solid lines, identical to Fig.~\ref{SpecificEmag}). Specific magnetic energies within the outer halo shell decrease over time as gas accretes. Note the contribution of outflows driven by stellar feedback to 
$\varepsilon_\text{mag}$ in the shell.}%
\label{sEmShells}%
\end{figure}

\subsection{Turbulent amplification: the accretion phase}
\label{ss:AccAmplification}

As previously mentioned, accretion of pristine gas from $t_\text{coll}$ onwards contributes to reducing specific magnetic energies due to its lower magnetisation. As shown in Fig.~\ref{sEmShells}, the specific magnetic energy measured within a thin spherical shell located at $r_\text{halo} < r < 1.05\, r_\text{halo}$ is significantly lower than the specific magnetic energy within the galactic region and exhibits a tendency to slowly decline with time. The presence of stellar feedback somewhat alleviates this decay by pushing a fraction of the magnetic field amplified in the galactic region out in the halo via outflows. This explains why both the B-Mech and B-RdTh runs show specific magnetic energies in the outer halo shell which are systematically larger than those of the B-NoFb run. Note that the specific magnetic energy of the virialized intra-halo region (i.e. $r < r_\text{halo}$) would therefore decrease if the halo were only to accrete this pristine, weakly-magnetised fresh gas.

In the accretion-dominated phase, extending from $z_\text{coll} \sim 13$ until $z_\text{acc} \sim 4$,  the galaxy continuously grows in mass and size, principally through accretion of cold filamentary flows \citep{Tillson15}. Once the collapse phase has ended and the average gas density of the galaxy is more or less fixed, growth of specific magnetic energy beyond $\varepsilon_\text{mag,C}$ requires stretching the magnetic field through dynamo processes. In particular, a turbulent ISM can generate exponential amplification of the magnetic energy through a turbulent dynamo
\citep{Rieder16}. The level of ISM turbulence is highly dependent on gas accretion, either directly \citep{Klessen10, Elmegreen09b} or through accretion-fuelled processes such as gravitational instabilities \citep{Elmegreen10, Krumholz16}. Even when turbulence is mainly attributed to stellar feedback processes, accretion is still credited with playing an increasingly relevant role, the higher the redshift becomes \citep{Hopkins13}. \citet{Klessen10} find that it mainly is the total mass of the galaxy that determines the importance of accretion-fuelled turbulence.  
The halo mass also sets whether cold accretion operates \citep{Birnboim03, Dekel06, Ocvirk08}. As galaxies and haloes accumulate mass, cold accretion progressively shuts down and the resulting turbulence rapidly decays.

In the absence of feedback, virialized hot halo gas can cool and fall onto the galaxy, but at the redshifts and halo masses discussed in this work,   feedback is required to maintain the presence of a significant hot gas phase \citep{Agertz09b} as accretion onto the galaxy proceeds mainly along cold filaments \citep{Tillson15}. However, feedback could also be reducing the budget of both absolute and specific energies available to drive turbulence through accretion by launching massive and disruptive galactic winds.

\begin{figure}%
\includegraphics[width=\columnwidth]{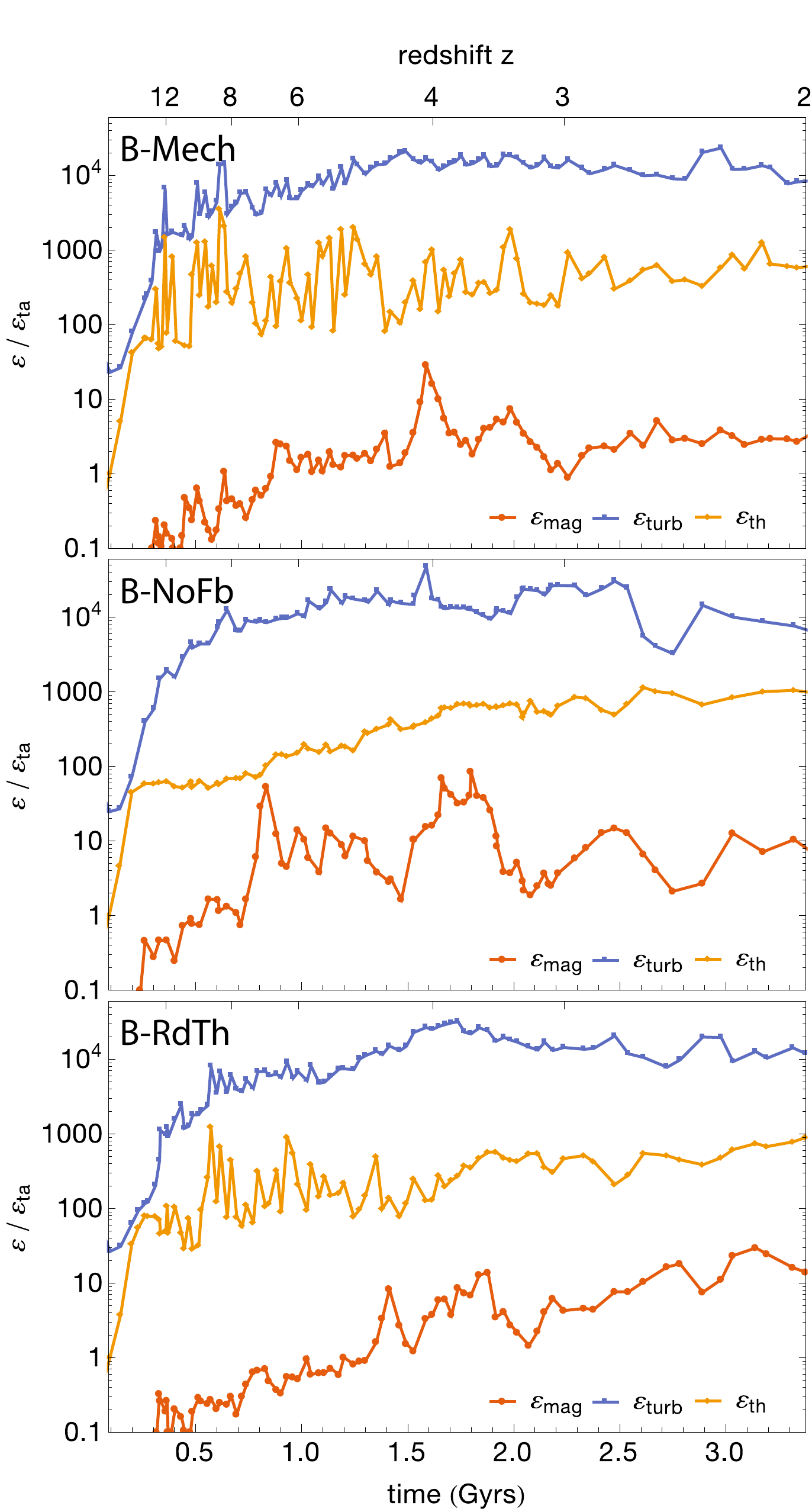}
\caption{Comparison of the specific magnetic (red), thermal (yellow) and turbulent (blue) energies  within the galactic region for the B-Mech (top), B-NoFb (middle), B-RdTh (bottom) runs. All energies have been normalized to a unique $\varepsilon_\text{ta}$ value (see text). To facilitate the comparison, $\varepsilon_\text{mag}$ curves have been boosted by 15 orders of magnitude. }%
\label{sEall}%
\end{figure}

\begin{figure*}%
\includegraphics[width=2\columnwidth]{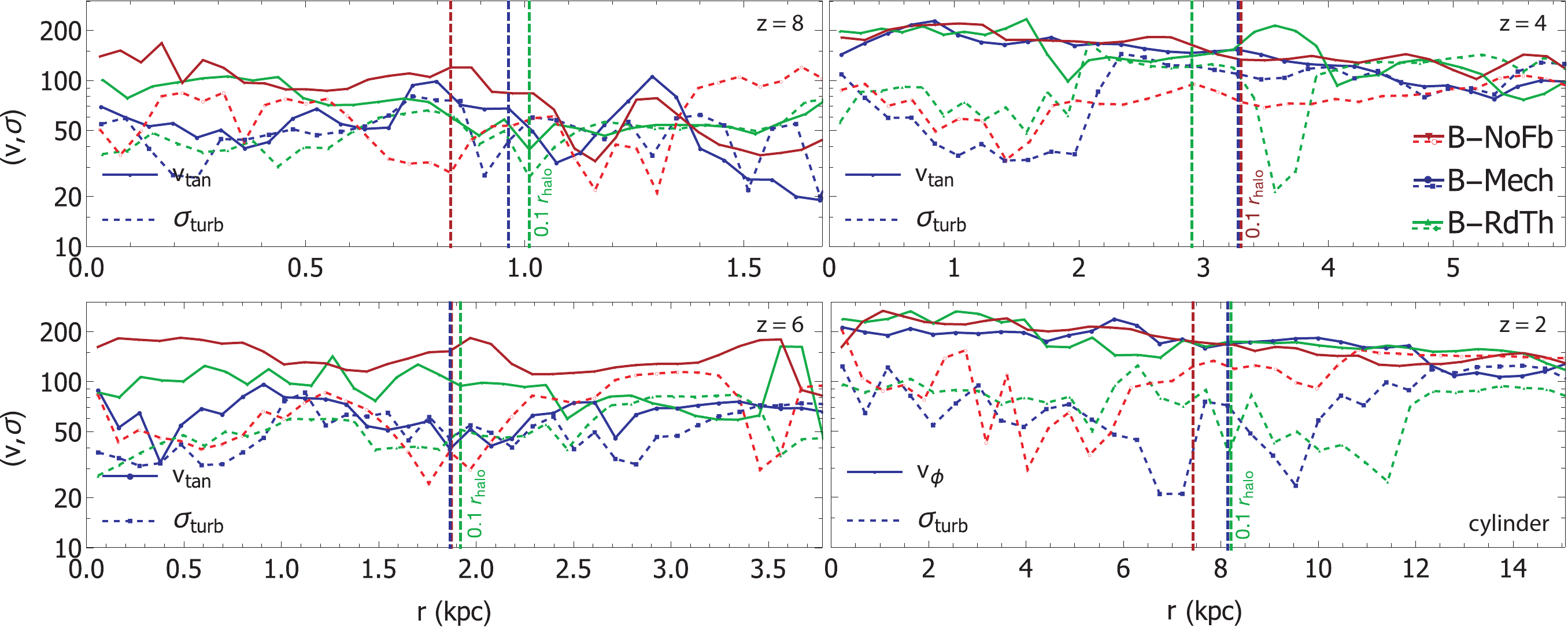}
\caption{Galactic region radially averaged velocity profiles for B-Mech (blue), B-NoFb (red), and B-RdTh (green). Cylindrical coordinates are employed for $z = 2$, with the poloidal component pointing in the direction of the galactic region angular momentum vector. Solid lines display the tangential, $v_\text{tan}$, (or toroidal, $v_\phi$,) component, dashed lines the turbulent velocity dispersions $\sigma_\text{turb}$. Vertical dashed lines represent $0.1\, r_\text{halo}$. As redshift decreases, galaxies become less dominated by turbulence and a clear rotation curve is established. }\label{rSProf}%
\end{figure*}

Bearing this possibly complex picture in mind, the evolution of turbulence within the galactic region is analysed 
by measuring in turn specific turbulent energies (Figure \ref{sEall}), radial velocity profiles (Figure \ref{rSProf}) and 
turbulent kinetic energy spectra (Figure \ref{Spectra}).

All specific energies displayed in Figure \ref{sEall} have been normalized to
$\varepsilon_\text{ta} = 10^{10} \text{cm}^2 \text{s}^{-2} $ 
which represents the specific thermal energy of the galactic region at turn-around. The turbulent energy $E_\text{turb}$ is calculated
by summing over the contribution of each cell enclosed in the galactic region:
\begin{equation}
E_\text{turb} =\sum_{i = 1}^{N_\text{cells}}{ \frac{1}{2} m_i  v^2_{\text{turb},i}},
\label{eq:ETurb}
\end{equation}
where $m_i$ is the gas mass, and the turbulent velocity of the gas cell is defined as  
$v_\text{turb,i} = \sqrt{v_{r,i}^2+(v_{t,i} - v_c (r))^2}$, with $v_{r,i}$ and $v_{t,i}$ the radial and tangential components of the gas velocity respectively, and the bulk motion of the galactic region has been subtracted. To remove the circular velocity $v_c(r) = \sqrt{GM(r)/r}$, where $M(r)$ is the total mass within the sphere of radius $r$, two distinct procedures are employed depending on whether a disk with substantial rotational support is present. If the measured spin parameter of the galactic region, $\lambda = L / (\sqrt{2} \, M(r_\text{gal}) \, v_c(r_\text{gal}) \, r_\text{gal}) < 0.5$ (with $L$ the angular momentum as in \cite{Bullock01}),  $v_c(r)$ is subtracted from the tangential velocity (i.e. the vector formed by the two non-radial components
of the velocity field) in each cell. After the formation of the disk, when $\lambda > 0.5$, $v_c$ is only subtracted from the toroidal velocity, in a cylindrical coordinate system where the $z$-axis is aligned with the gas angular momentum of the disk and goes through the centre of the galactic region. Turbulent velocity dispersions profiles, $\sigma_\text{turb}(r)$, built on this definition of $v_\text{turb,i}$, are presented in Fig. \ref{rSProf}. The formation of a stable gas disk occurs around $z \sim 4.5 \text{-} 4.0$ in all three simulations, and is confirmed by visual inspection. This transition correlates with the end of the accretion phase and occurs shortly after the gravitational potential of the galactic region becomes dominated by the stellar component. For simplicity of the analysis, the transition is taken to occur at the same $t_\text{acc/fb}$ ($z_\text{acc/fb} = 4$) in all runs.  

Two features are identified during the accretion-dominated phase in Fig.~\ref{SpecificEmag}: (i) sharp peaks are present throughout the entire phase, and (ii) the underlying continuum upon which these peaks rest grows steadily with time. We attribute the presence of peaks to individual mergers, while the growth of the smooth underlying component corresponds to {\em in situ} turbulent amplification. They are consequently addressed separately.

Monotonic turbulent amplification of the specific magnetic energy results in an exponential growth $\varepsilon_\text{mag} \propto e ^{\Gamma_{\varepsilon} t}$, with characteristic time scale $\Gamma_{\varepsilon}$. Such an exponential growth typically indicates a turbulent dynamo process \citep{Federrath16}. Exact values of characteristic time scales obtained by fitting exponential curves to $\varepsilon_\text{mag}$ (see Fig. \ref{SpecificEmag}) are given in Table \ref{table:EmagGrowth} for both accretion and feedback-dominated epochs.
During the accretion dominated epoch, they are approximately on the order of $\Gamma_{\varepsilon, \text{acc}} \sim 2 \text{ Gyrs}^{-1}$, 
in fair agreement with the analytic estimate $\Gamma_\varepsilon \sim \sigma_\text{turb} / l \sim 1.5\, -\, 3 \text{ Gyrs}^{-1}$, provided one picks a
dissipation scale, $l \sim 10\, \text{-} \, 20$ pc, typical of the maximal resolution achieved in our AMR simulations.  
From the numbers listed in Table \ref{table:EmagGrowth}, we see that during the accretion phase, $\Gamma_{\varepsilon}$ is larger in the absence of feedback. The amplification is also reduced as the strength of the sub-grid feedback implementation increases (the B-RdTh run yields a lower value than the B-Mech run). Stellar feedback thus has a detrimental effect on magnetic energy amplification during the accretion phase. Potential explanations of this behaviour are an increase in turbulent diffusion caused by stellar feedback \citep{Gressel08}, a suppression of gravitational instabilities, feedback diluting magnetic fields into the hot phase, or a reduction of small scale solenoidal turbulence relative importance \citep{Grisdale17}. More specifically, stellar feedback could be hampering clump formation triggered by larger scale gravitational instabilities in the galaxy. As solenoidal modes are expected to produce higher growth rates \citep{Federrath11b} and stellar feedback can efficiently drive solenoidal turbulence \citep{Korpi99} one would expect that in the presence of stellar feedback, the fraction of specific turbulent energy in solenoidal modes within the galactic region should increase. However, we find the opposite during the accretion phase: stellar feedback reduces the fraction of solenoidal specific turbulent energy in our runs on scales $L < 1 \text{kpc}$ by $\sim 10\%$. A detailed study of this complex interaction between feedback, turbulence modes, and magnetic fields amplification is beyond the scope of this paper. However we note that, despite the anisotropic nature of the stellar feedback implementation in the B-Mech run, which has the potential to yield a higher amplification of the magnetic energy than the B-NoFb run because of an intrinsically more efficient capability to drive solenoidal turbulence than its isotropic B-RdTh counterpart, this does not happen in our simulations. This interpretation is underpinned by the measurement of the specific turbulent energies presented in Fig~\ref{sEall}. It is clear from Fig~\ref{sEall} that, at least for $z >4$, the level of turbulent energy (blue curves) measured in the B-NoFb run is at least comparable to, if not higher than that measured in both feedback runs. We point out that these findings corroborate both the observational standpoint that stellar feedback does not rank as the favoured mechanism to drive turbulence in high-redshift clumpy galaxies \citep[e.g.][]{Elmegreen09a} and cosmological simulations of magnetised Milky Way-like galaxies performed with a different (moving mesh + divergence cleaning) technique \citep{Pakmor14}.

\begin{figure*}%
\includegraphics[width=2.\columnwidth]{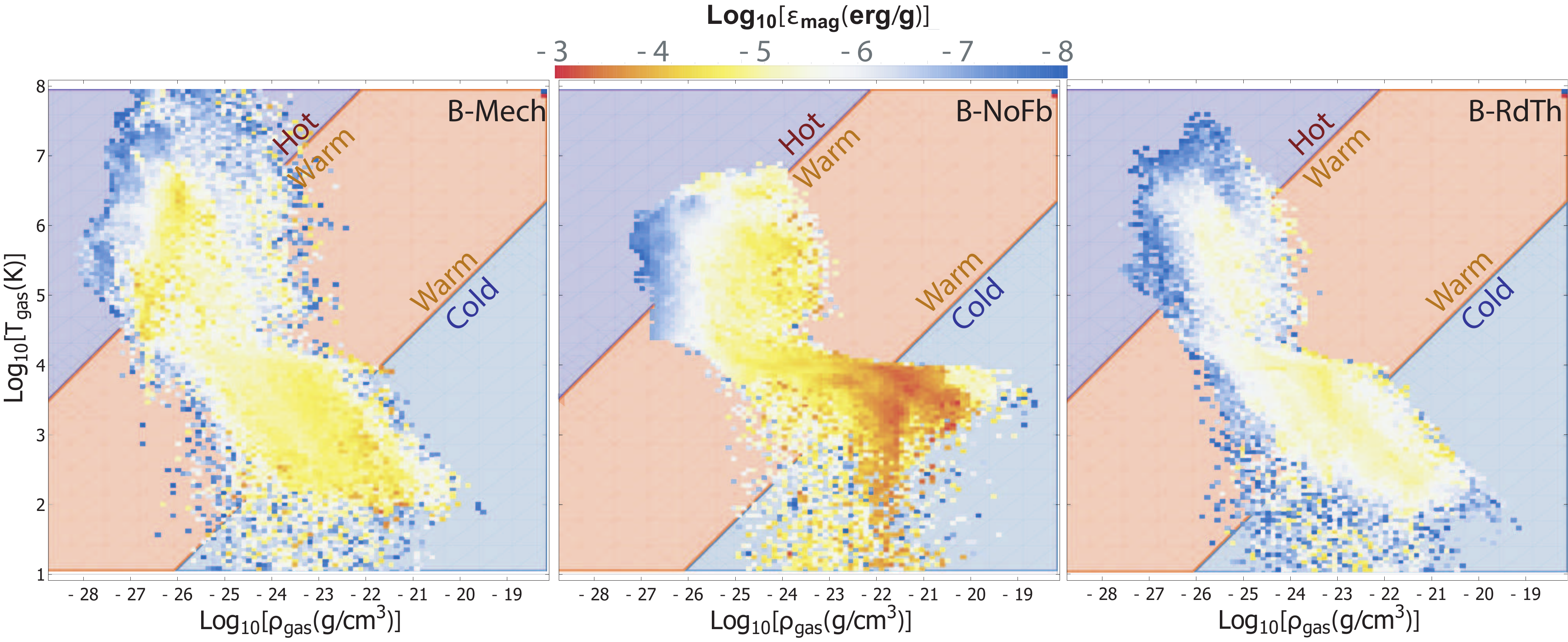}
\caption{Gas density ($\rho_\text{gas}$) - temperature ($T_\text{gas}$) phase diagrams colour coded by specific magnetic energy ($\varepsilon_\text{mag}$). Diagrams correspond to the galactic regions at $z = 6$. Simulations from left to right are B-Mech, B-NoFb and B-RdTh. Following \citet{Gent12}, three ISM phases are identified by shaded regions: hot (purple, top left), warm (red, central), and cold phases (blue, bottom right). Divisions between ISM phases correspond to lines of constant specific entropy $s \propto \left[\log{T} - \left(\gamma - 1\right) \log{\rho_\text{gas}}\right]$.}%
\label{PhaseDiags}%
\end{figure*}

\begin{figure}%
\includegraphics[width=\columnwidth]{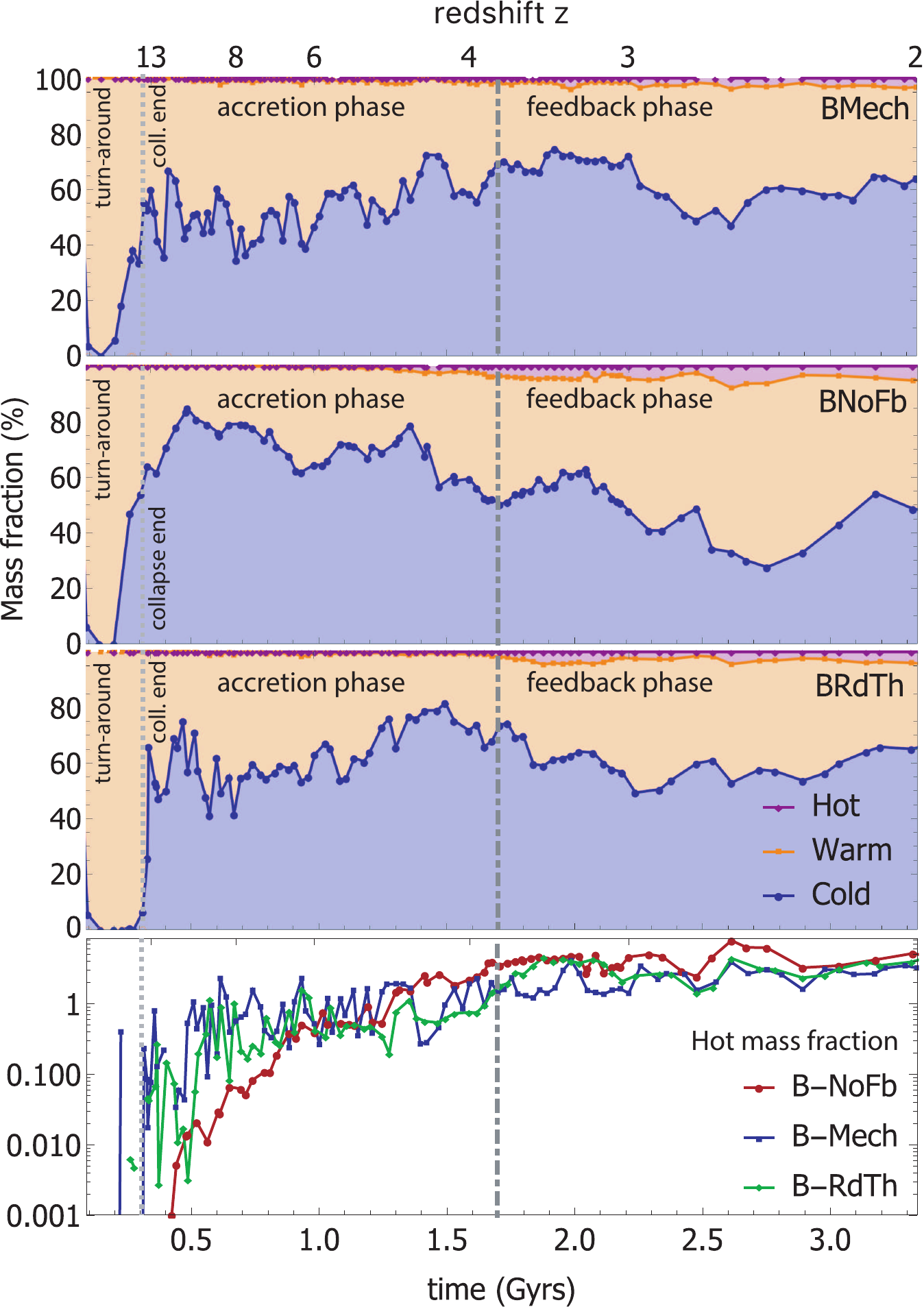}
\caption{Gas mass fraction within the different ISM phases of the galactic region (hot - purple; warm - orange; cold - blue). Runs are from top to bottom B-Mech, B-NoFb, and B-RdTh. For clarity, the bottom panel presents a zoom version of the hot phase mass fractions for the three runs.}%
\label{coldFraction}%
\end{figure}

As feedback heats the ISM gas, another significant difference between the B-NoFb run and its feedback counterparts, is that this former displays a larger difference between thermal and turbulent specific energies (yellow and blue curves in Fig.~\ref{sEall} respectively). This supports the claim by \citet{Evirgen17} that the contribution from the hot phase of the ISM to the amplification of magnetic energy is negligible. A possible explanation may be that flow resides in this phase for a short time compared to the dynamo time scale. By contrast, these two timescales become comparable for the warm and cold phases, indicating that these latter are responsible for most of the amplification. For these reasons, we expect the cold phase to be the most efficient at amplifying the field in our simulations. In an effort to quantify this, Fig.~\ref{PhaseDiags} displays density-temperature ($\rho_\text{gas} \text{-} T_\text{gas}$) phase diagrams, colour coded by specific magnetic energy, for the galactic region at $z=6$. In the three runs, we have divided the ISM following \citet{Gent12,Evirgen17}. We distinguish three different ISM phases: cold and dense, warm, and hot and diffuse. Hereafter we will simply refer to them as cold (blue), warm (orange), and hot (purple). These phases are separated by constant specific entropy lines in Fig.~\ref{PhaseDiags}. The fraction of gas mass contained in each phase is shown in Fig.~ \ref{coldFraction}. We find that the fraction of gas in the hot phase is negligible (less than 10\%) in all three runs and at all times (see bottom panel of Fig.~\ref{coldFraction}). During the very early stages of evolution ($z \geq 7$), the amount of hot gas in the B-NoFb run is significantly lower than that of both feedback runs \citep{Agertz09a}, but it progressively grows as the the cooling time scale overtakes the compression time scale and the stable standing shock is pushed outside of the galactic region \citep{Dekel06}. Indeed, at later times ($z \leq 4.5$),
the B-NoFb run even has comparatively more hot gas than both feedback runs due to its less efficient, quasi metal-free cooling \footnote{Note however that, at least for $z > 2.5$, this hot halo gas is unable to sever the filaments of cold gas directly supplying the galactic region \cite{Tillson15}.}.
In any case, as is clear from Fig~\ref{PhaseDiags} and Fig~\ref{coldFraction}, most of the specific magnetic energy concentrates in the mass dominant cold phase as well as, to a lesser extent, the colder part of the warm phase. This trend is especially marked for the no B-NoFb run (middle panel of Fig~\ref{PhaseDiags}) where $\varepsilon_\text{mag}$ is strongly concentrated in the density and temperature ranges  $10^{-23} \text{g} \, \text{cm}^{-3} < \rho_\text{gas} < 10^{-20} \text{g} \, \text{cm}^{-3}$ and $10^3 \, \text{K} < T < 10^4 \, \text{K}$ respectively. In the feedback runs, the specific magnetic energy occupies a wider locus in the phase diagram, and is more homogeneously distributed across the cold and warm phases. As previously mentioned, this reinforces the idea that SN feedback is detrimental to magnetic energy amplification during the accretion phase: because feedback heats up the gas, during most of the accretion-dominated phase ($12 > z > 4.5$), the B-NoFb run contains on average $10 - 20 \%$ more gas in the cold phase than both feedback runs (Fig.~\ref{coldFraction}). 

Another striking feature of the phase diagrams presented in Fig.~\ref{PhaseDiags} is the absence of correlation between specific 
magnetic energy and gas density in all three runs, which would have, had it been present, indicated an adiabatic-like amplification 
of the magnetic field. Instead, Fig~\ref{rSProf} shows that the B-NoFb run has the largest absolute velocity dispersion $\sigma_\text{turb}$ 
of the three simulations throughout a large fraction of the galactic region ($r \lesssim 0.1 \; r_\text{halo}$) at high redshift ($ z \geq 4.5$). 
This is not a very large effect, as these values are generally comprised between 40 - 80 km s$^{-1}$ in the B-NoFb run, compared to 30 - 60 km s$^{-1}$ for the B-Mech and B-RdTh runs, but it is nonetheless systematic. In all three runs, the turbulent velocity profiles subtly increase with $r$, so that in the external galactic region ($r \gtrsim 0.1 \; r_\text{halo}$), they all reach typical values on the order of $\sigma_\text{turb} \sim 50 \text{-} 100$ km s$^{-1}$, mainly dominated by radial velocity dispersion. This is not surprising as turbulence is expected to be higher in the external region of galaxies when significant accretion flows are present \citep{Klessen10, Hopkins13}. As redshift decreases, the velocity dispersion in the B-RdTh and B-Mech runs increases monotonically, to reach a similar (or even slightly higher) level that in the B-NoFb run by $z = 4$. It is interesting to note it is at this redshift when the disk forms, even though there are hints of an earlier presence in the B-NoFb run. 

\begin{table}
\centering
\caption{$\Gamma_\varepsilon$ timescales derived from exponential fits to specific magnetic energies in the accretion phase, 
$\Gamma_{\varepsilon, \text{acc}}$, and the feedback phase, $\Gamma_{\varepsilon, \text{fb}}$. All $\Gamma_{\varepsilon}$ time-scales are measured in Gyrs$^{-1}$. See text (section \ref{s:Robust}) for details of the fitting procedure.}
\label{table:EmagGrowth}
\begin{tabular}{l | r r}
\hline
Run & $\Gamma_{\varepsilon,\text{acc}}$ & $\Gamma_{\varepsilon,\text{fb}}$\\
\hline
B-NoFb & $2.7 \pm 0.2$ & $-2.0 \pm 0.5$ \\
B-Mech & $2.3 \pm 0.1$ & $0.3 \pm 0.2$ \\
B-RdTh & $2.0 \pm 0.1$ & $1.7 \pm 0.2$ \\
\hline
\end{tabular}
\end{table}

Throughout the accretion period and particularly at the end ($z \sim 4$), all three runs display sharp peaks in their specific magnetic energies (see Fig.~\ref{SpecificEmag}). These peaks are correlated with mergers, which stretch the magnetic field out of galaxies in tidal streams during close encounters, as reported in previous studies \citep[see][]{Roettiger99,Kotarba11}, and are followed by a subsequent decrease in magnetic energy generally attributed to numerical dissipation \citep{Kotarba11,Geng12,Pakmor14}. Note that the main difference between our work 
and these simulations (besides our order of magnitude better spatial and mass resolution which makes the merger peaks more prominent) is that they typically reach magnetic field values close to saturation before the merger happens, regardless of the initial value of the magnetic field they use. This does not happen in our simulations with the magnetic seed employed here. We do reach energy equipartition for higher initial values of the magnetic field (Martin-Alvarez et al. in preparation). Such a saturation of the magnetic field is generally interpreted in these papers as evidence of numerical convergence. However, given the high level of residual magnetic divergence inherent to some of the divergence cleaning methods used in these works, one is left to wonder about the role played by numerics in establishing the mean level of field amplification. This is especially worrisome given the results reported in the comparative study presented by \citet{Mocz16}, where the magnetic pressure in two idealised isolated disk simulations that differ only by their induction equation solver (CT versus divergence cleaning) 'converges' to values different by more than an order of magnitude (see Fig. 6 of \citet{Mocz16}). 

\begin{figure*}%
\includegraphics[width=2.\columnwidth]{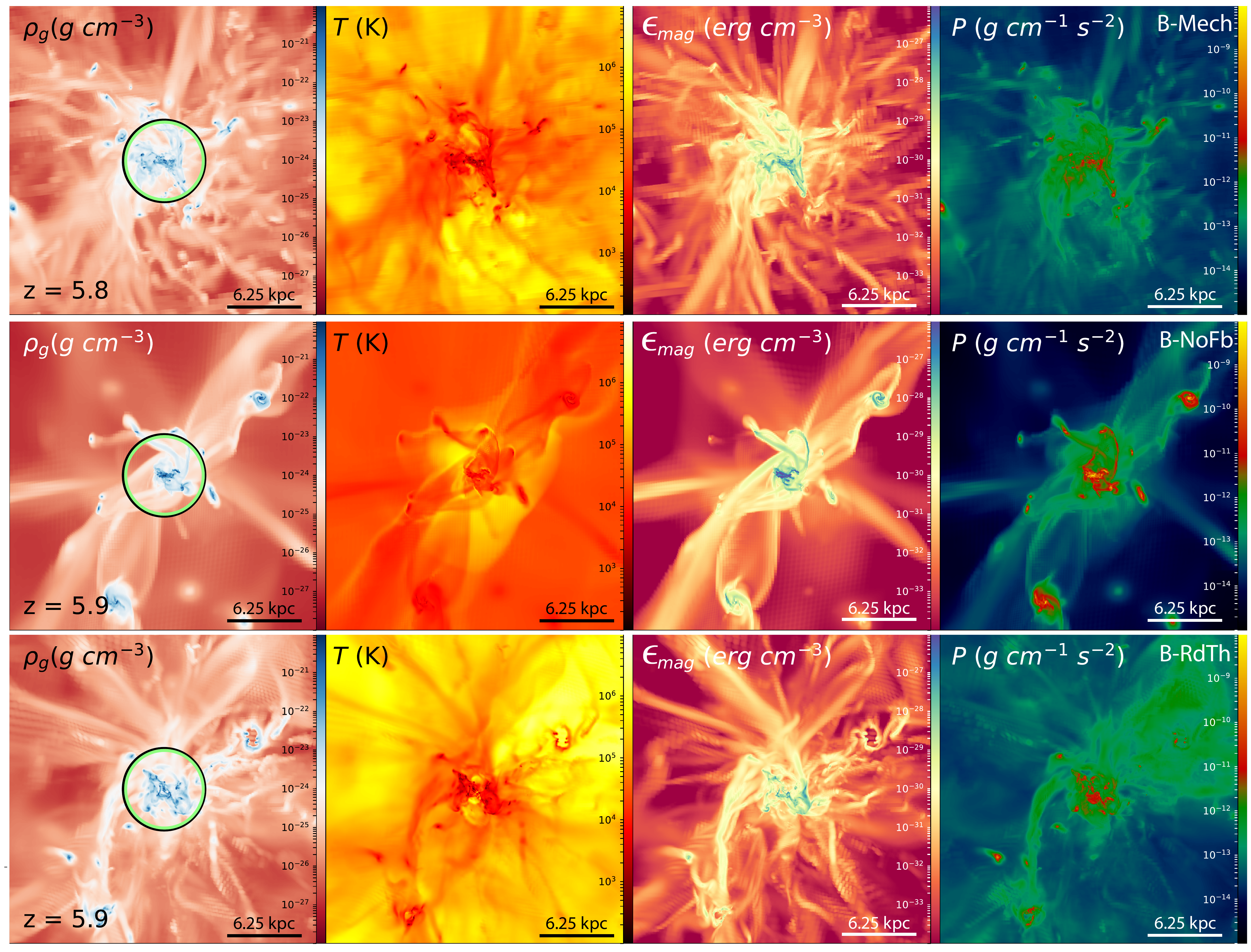}
\caption{Projections of a (25 kpc)$^3$ cubic volume centred on each galactic region for the three runs during a merger even ($z \sim 6$). Density, temperature, magnetic energy density and thermal pressure are displayed  in columns from left to right respectively (B-Mech run: top panels; B-NoFB run: centre panels; B-RdTh run: bottom panels). In both feedback runs magnetic field lines around the central galaxy are disrupted by the strong outflows of hot gas. These outflows reduce the efficiency of magnetic stretching resulting from the merger, 
generating shallower peaks in the specific magnetic energy compared to the no feedback run. Galactic regions are indicated as circles on the gas density images (left column) and intensive variables are mass weighted. The exact redshifts for each image correspond to the points circled in black on Figure \ref{SpecificEmag}.}%
\label{MergerImages}%
\end{figure*}

\begin{figure*}%
\includegraphics[width=2\columnwidth]{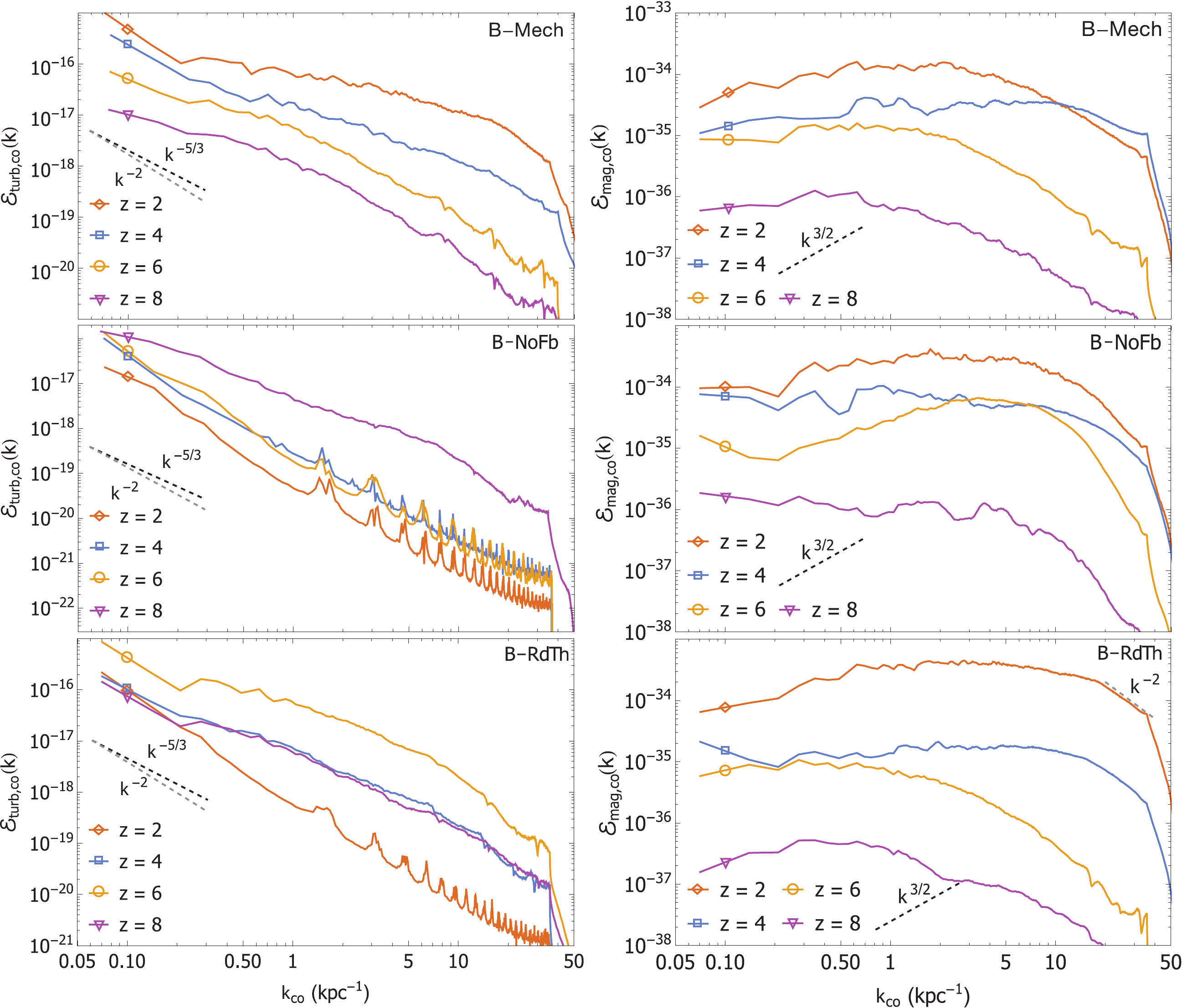}
\caption{Comoving turbulent kinetic (left) and magnetic (right) energy power spectra for B-Mech (top row), B-NoFb (middle row), and B-RdTh (bottom row). Dashed lines represent Kolmogorov (black), and Burgers (gray) turbulent scalings for the kinetic energy; and Kazantsev scaling 
for the magnetic energy. Spikes in the turbulent kinetic energy spectra correspond to resonances with nested AMR grid levels.}%
\label{Spectra}%
\end{figure*}

Fig.~\ref{MergerImages} displays projections of the gas density, temperature and pressure, along with the specific magnetic energy in a volume  centred on the galactic region and coincident with a merger occurring around $z \sim 6$ in all three runs. The peaks in the specific magnetic energy coincident with the merger event are also marked as black circles in Fig~\ref{SpecificEmag}. From the combination of Figs~\ref{SpecificEmag} and 
\ref{MergerImages} is clear that mergers boost magnetic energy within the galactic region (solid circles on the gas density panels of Fig~\ref{MergerImages}) by stretching field lines. Moreover, $\varepsilon_\text{mag}$ peaks in the B-NoFb run achieve considerably higher values  than their feedback run counterparts, as field lines smoothly follow merger stream flows in the absence of feedback. In the presence of feedback these very same field lines are strongly disrupted, and even though the level of turbulence may rise within the galactic region and amplify the field, a fraction of the resulting magnetic energy is expelled by outflows into the halo \citep[see also][]{Dubois10}, generating the radial filamentary magnetic structures seen in the magnetic energy density maps of Fig.~\ref{MergerImages}. More quantitatively, a higher specific magnetic energy is measured in the outer regions of halos in the feedback runs (see Fig.~\ref{sEmShells}). As a consequence, even though a significant amount of magnetic energy is generated by bulk stretching during mergers, a large fraction of this energy is lost through outflows in the presence of feedback. Once the merger concludes and the galaxy settles, significant numerical magnetic reconnection occurs as the gas (and field) falls onto the galaxy and further of the magnetic energy gained in the merger is lost (see Figs~\ref{Resolution} and \ref{ResolutionPhases} where the amplitude of the merger peaks is greatly reduced in lower spatial resolution, but otherwise identical runs). Since these losses are strongly correlated with resolution, better resolving the circum-galactic medium seems required in order to properly capture the amplification of magnetic energy by mergers. At fixed resolution, the use of fixed nested grids as implemented in \citet{Vazza14} for LSS simulations, which set a unique turbulent and magnetic energy dissipation scale, also alleviate the problem. Bearing in mind these resolution caveats,  we find, in agreement with \cite{Rieder17b} for cosmological simulations of dwarf galaxies, that mergers have a negligible contribution to the evolution of the magnetic energy within the galactic halo.
 
To further assess the presence of a turbulent dynamo, we measure turbulent kinetic and magnetic energy spectra for each simulation (Fig.\ref{Spectra}). These are calculated by performing a Fast Fourier Transform (FFT) of cubic boxes with $1024^3$ cells, centred on the galactic region in each run, and with physical box length of $l = 0.4\,r_\text{halo}$, so as to encompass the whole galactic region. For the sake of numerical performance, our FFT calculation assumes periodic boundary conditions, which, although obviously unrealistic at these scales, still provides a very good approximation  because of the high gas density and magnetic field contrast between the galaxy and its surroundings. In other words, the tenuous circumgalactic medium acts as a zero-padding region which isolates the galaxy, ensuring that the power in Fourier modes at any scale but the largest is 
dominated by that of the galaxy ISM. The impact of periodicity and shape of the galaxy disk on the measured spectra are discussed in more detail in Appendix \ref{ap:diskSpec}, so we only mention at this stage that the disk shape results in a power law decay of the density spectra as a function of the spatial angular wavenumber $\propto k^{-7/4}$ for $k_s \geq \frac{\pi}{h_s}$. As previously discussed, kinetic energies were calculated from velocities with bulk motion and circular velocity $v_c (r)$ removed. All spectra are displayed in comoving units: dark matter halo physical radii grow $\propto (1+z)^{-1}$, so their comoving size remains roughly constant with time, which facilitates the analysis and comparison of the different spectra across the entire redshift range.

The turbulent kinetic energy spectra for B-NoFb from $z \sim 8$ to $z \sim 4$ exhibit a cascade with power law scaling between Kolmogorov $\mathcal{E}_\text{turb} \propto k ^{-5/3}$ \citep[][indicated by black dashed lines in the left column panels of Fig. \ref{Spectra}]{Kolmogorov41} and Burgers $\mathcal{E}_\text{turb} \propto k ^{-2}$ \citep[][indicated by gray dashed lines on the same panels]{Burgers48} at intermediate scales $ 0.3 \text{ kpc}^{-1} < k_\text{co} < 3 \text{ kpc}^{-1}$, i.e. in the inertial range, where $k_\text{co}$ represents the co-moving spatial angular wavenumber. These cascade spectra for the turbulent kinetic energy indicate a transfer of energy from large scales towards smaller eddies, consistent with fully developed turbulence resulting from continuous accretion. On the other hand, the feedback runs, and especially B-Mech, inject energy and/or momentum on small scales \citep[see e.g.][for detail]{Kimm15}, as expected from SNe explosions. This generates extra power on scales ranging typically from $3 \text{ kpc}^{-1} < k_\text{co} < 30 \text{ kpc}^{-1}$, flattening the power law index of the spectra as the scale height of the disk increases (see Appendix \ref{ap:diskSpec}). Note that the energy injection scale into the galaxy can be identified by the break present in all spectra around comoving wavenumber $k_\text{co} \sim 0.2 \text{ kpc}^{-1}$ (corresponding to physical wavelengths of $\sim$3 and $\sim$10 kpc at redshifts 8 and 2 respectively; see left column panels of Fig~\ref{Spectra}). 

Because small scale turbulence more rapidly amplifies the magnetic field due to its short e-folding times, the magnetic energy generated by the dynamo is fed back towards large scales, as evidenced by an inverse cascade in the magnetic energy spectra (right column panels of Fig.~\ref{Spectra}). In agreement with simulations by \citet{Bhat13, Federrath16} these inverse cascade spectra typically have a shallower slope than the Kazantsev spectrum \citep[$\mathcal{E}_\text{mag}\propto k^{3/2}$]{Kazantsev68} on large scales ($k_\text{co} \leq 1 \text{ kpc}^{-1}$), characteristic of compressible flows with magnetic Prandtl numbers $\text{Pm} > 1$ in the kinematic stage (the magnetic energy is much smaller than the kinetic energy see Fig.\ref{sEall}). We estimate the numerical $2 < \text{Pm} < 10$ for our ideal MHD scheme, see \cite{Rembiasz17} for detail. We argue that part of this shallowness arises from the disk-like shape of our galaxies and the complexity of the system (bursty injection of energy and momentum by SN in a multiphase ISM subject to accretion and mergers). Indeed, when going from the largest (box size) to the smallest scales (single grid cell), the power in magnetic energy spectra of isothermal turbulent boxes first increases as steeply as Kazantsev, then flattens as it reaches a maximum on scales corresponding to multiple elements of resolution, before decaying \citep[e.g. Fig. 5 of][]{Haugen04}. Our simulations display the same global qualitative behaviour but the inverse cascade spans a range of wave numbers which comprises both the scale length and scale height of the galaxy, so the galaxy shape naturally influences the energy spectra slopes. A perhaps clearer imprint of this effect can be found on the smallest scales (large $k$s) where our spectra exhibit a steep 'disk-like' decay following $\mathcal{E}_\text{mag} \propto k^{-7/4}$ (see Fig. \ref{DiskDecay} and discussion in the appendix \ref{ap:diskSpec}) instead of the somewhat shallower decay observed in simulations of turbulent boxes \citep[both for compressible and incompressible fluids]{Haugen04, Schekochihin04, Bhat13, Federrath16}. Isolated galaxy simulations are expected to have less power on large scales (greater than 
the galaxy length scale) by construction. However, similar magnetic spectra to ours are also observed in such simulations \citep{Rieder16,Butsky17}, and in a cosmological simulation of a dwarf \citep{Rieder17b}. A difference worth noting is that \citet{Butsky17} (see their Fig. 5) obtain magnetic energy spectra in the kinematic regime steeper than Kazantsev on intermediate to large scales, which we speculate might be the result of their SN seeding of magnetic field.

\begin{figure}%
\includegraphics[width=\columnwidth]{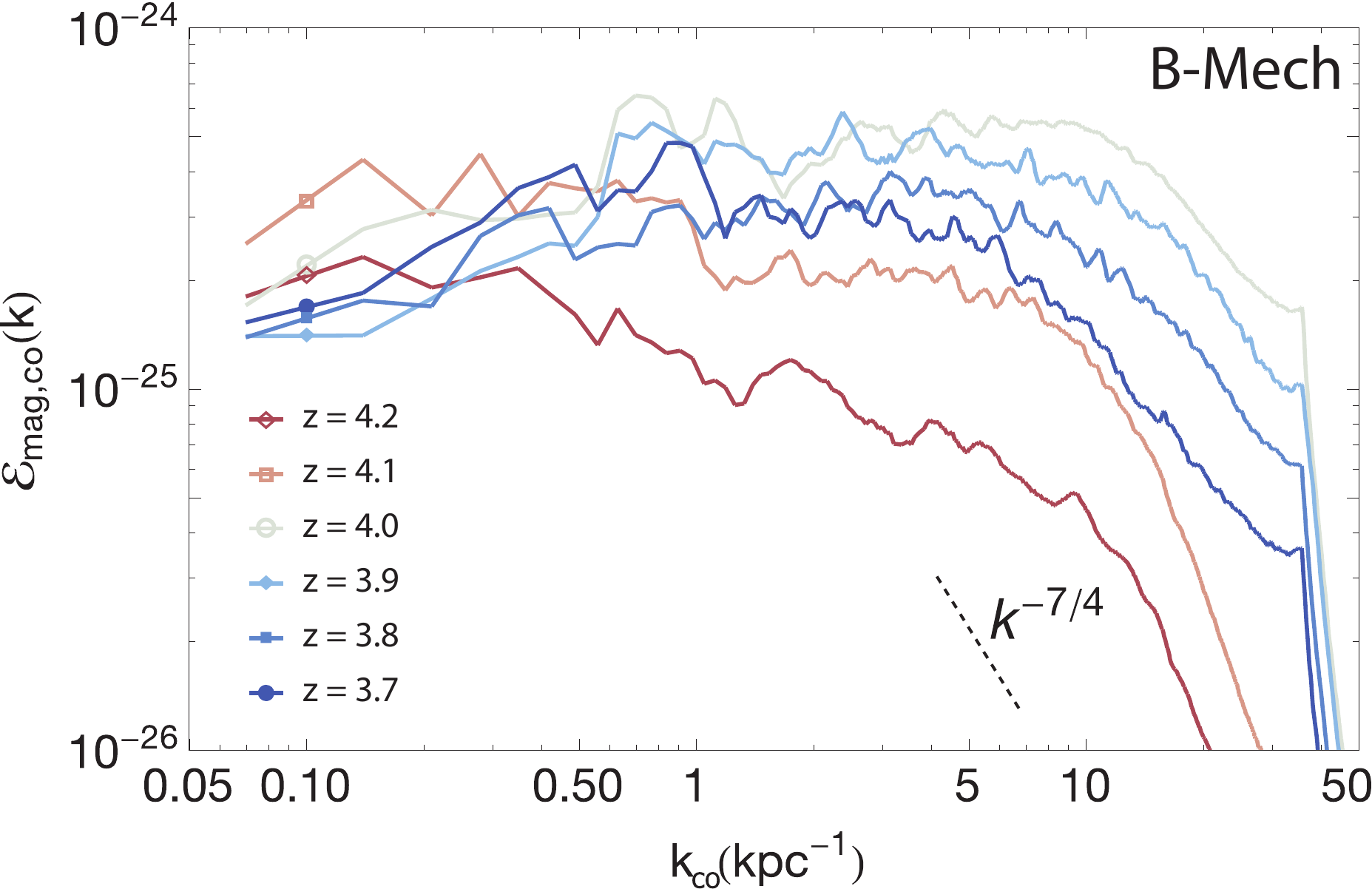}
\caption{B-Mech comoving magnetic energy spectrum evolution throughout the merger occurring around $z = 4$. Redshift evolves from high (red) to low values (blue). As the merger proceeds, power is gained on small scales due to the window function set by the disk being blurred. Once the merger has finished, the field is confined in the galaxy again and the decay due to the shape of the disk progressively propagates to larger scales.}%
\label{ContSpec}%
\end{figure}

For a more detailed look at the time evolution of the magnetic energy spectra when a merger occurs, Figure \ref{ContSpec} displays the evolution of $\mathcal{E}_\text{mag}$ for B-Mech, around a merger event at $z = 4$. As redshift decreases, the stretching resulting from material and magnetic field being pulled out of the galaxy amplifies the field. Furthermore, as the magnetic field is stretched, the window defined by the disk of the galaxy is temporarily blurred, eliminating the the $k^{-7/4}$ decay, and the spectrum displays an inverse cascade all the way to $k_\text{co} \sim 10 \text{ kpc}^{-1}$ ($k_\text{ph} \sim 50 \text{ kpc}^{-1}$). However, once the merger completes, tidal gas streams fall back onto the galaxy and magnetic energy is lost due to numerical reconnection. Present day simulations seem to the lack resolution to properly capture the amplification of turbulence and magnetic energy associated with the merger process. As a result, mergers only act as temporary enhancers of the field, both in the galaxy and the circumgalactic medium \citep[see also][]{Rieder17b}.

The presence of a turbulent cascade in the kinetic turbulent spectra and an inverse cascade in the magnetic energy spectra between $z = 8$ and $z = 4$, 
regardless of whether SN feedback is implemented or not, proves that during the accretion phase, a turbulent galactic dynamo is triggered, amplifying the magnetic energy on small scales and feeding it back to larger galactic scales.

\subsection{Feedback driven amplification}
\label{ss:FbAmplification}

\begin{figure}%
\includegraphics[width=\columnwidth]{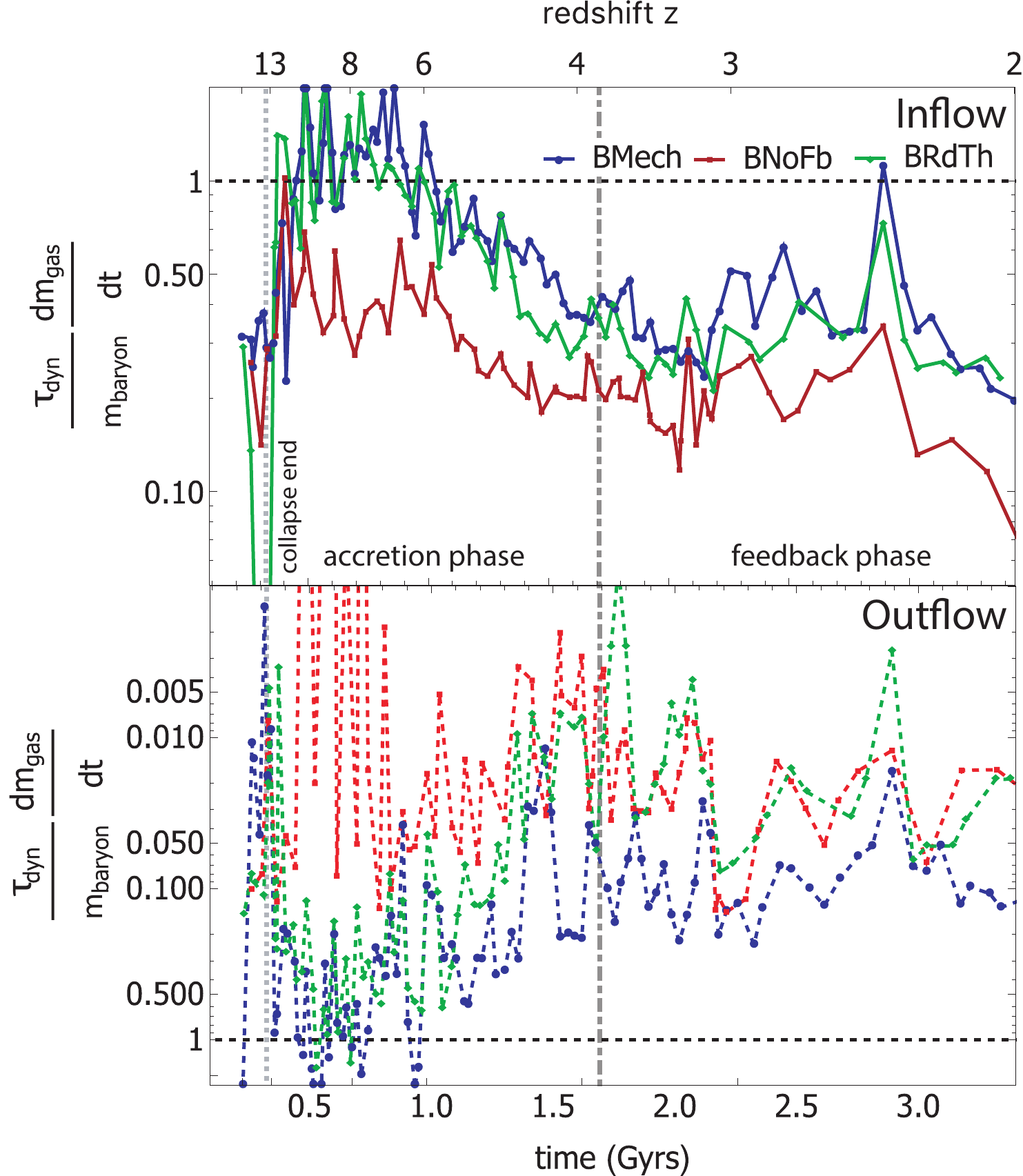}
\caption{Ratio of the inflowing (top) and outflowing (bottom) gas across the galactic region to the total baryonic mass contained within it during a dynamical time for all three studied runs. Dynamical timescales of the galactic regions are computed for each run as $\tau_\text{dyn} (z) = 2\pi\;0.2\;r_\text{gal} (z)\;v^{-1}_c\left( r_\text{gal} (z)\right)$.}%
\label{AccRate}%
\end{figure}

The final period of amplification is dominated by local galactic effects instead of accretion. Galaxies have grown to a size where gas accretion represents a small fraction of their baryonic mass content, as shown on Fig.~\ref{AccRate}. The progressive shut down of cold accretion flows, and the increased stability of the gas disk as the stellar component grows also reduce the impact of accretion \citep{Dekel06,Klessen10,Cacciato12}. Thus, inflow-induced turbulence declines, especially in the inner regions of the disks. Outer regions still seem to display accretion-driven turbulence \citep[see Fig.~\ref{rSProf} and][]{Klessen10, Hopkins13} that could maintain or even amplify the field locally \citep{Pakmor17}. Transfer of magnetic energy from the outer parts of galaxies towards the centre has been reported in other studies \citep{Machida13}, significant enough to strengthen the inner field once a dynamo powerful enough is activated in the external region. However, this does not happen in the simulations presented in this work. Instead,
as there are no sub-grid processes forcing turbulence nor injecting energy or momentum in the B-NoFb run, and accretion loses importance, both specific and absolute turbulent energies, required to maintain the growth of the field, decay (Table \ref{table:EmagGrowth}). Numerical resistivity also contributes to the decay of the magnetic energy, but the decline of turbulence eventually leads to the onset of the final field topology as argued in \cite{Rieder17a}. Note however that, contrary to what we measure for the B-NoFb run, even without an explicit feedback model \citep[that of][]{Springel03}, \citet{Pakmor14} report that after the formation of a Milky Way-like galaxy they still find linear growth of the magnetic field, which they attribute to differential rotation.

On the other hand, even after accretion has significantly diminished, runs with stellar feedback still exhibit considerable amounts of turbulent energy
throughout the galaxy \citep[at least an order of magnitude more that B-NoFb, see Fig. \ref{Spectra} and e.g.][for a discussion of stellar feedback-induced turbulence]{Federrath16}. It is now well established that compressive forcing due to SNe can drive solenoidal turbulence \citep{Korpi99,Mee06,Federrath10}, and considerable amplification is expected in trailing SN shock waves \citep{Suoqing16}. Having said that, because of considerable numerical turbulent and magnetic dissipation due to limited resolution, and the neglect of magnetic field injection by SN which might constitute a considerable source of magnetic energy \citep[see e.g.][]{Butsky17}, our results should be viewed as only providing a lower limit on the importance of SN during the feedback dominated phase. 

Bearing in mind these caveats, SN driven turbulence still causes an exponential amplification of the magnetic energy in runs B-Mech and B-RdTh through a turbulent dynamo (as indicated by the turbulent kinetic and magnetic energy spectra corresponding to this phase on Fig. \ref{Spectra}). Table \ref{table:EmagGrowth} shows that  B-RdTh displays a larger growth rate, $\Gamma_{\varepsilon_\text{mag}}$, than B-Mech. However, both runs have lower $\Gamma_{\varepsilon_\text{mag}}$ than during their accretion-driven phase. We suspect this is related to the fact that accretion and gravitational instabilities in a differentially rotating disk preferentially drive solenoidal turbulence which is more efficient at amplifying the magnetic field. Admittedly, this latter property will depend both on the exact subgrid implementation of stellar feedback and on resolution, as vorticity generated behind shocks is difficult to capture, but compressible turbulence conversion into solenoidal modes within the small volume covered by a 2D-like galaxy is limited. As B-RdTh is more efficient at perturbing the gaseous disk of the galaxy and driving a fountain, it is therefore not surprising that this run displays a higher $\Gamma_{\varepsilon_\text{mag}}$ than its B-Mech counterpart. B-Mech is in turn more efficient driving outflows (see Fig.\ref{AccRate}), which are responsible for injecting magnetic energy in the rest of the halo, causing the higher specific magnetic energies seen in Fig \ref{sEmShells}.
In summary, while in the accretion driven phase, feedback, if anything, is detrimental to amplification ($\Gamma_{\varepsilon_\text{mag}}$ values in Table \ref{table:EmagGrowth} are slightly lower for B-Mech and B-RdTh than for B-NoFb), in the feedback-driven phase it is the main driver of the amplification the field, which results in the values of $\Gamma_{\varepsilon_\text{mag}}$ being both very model-dependent during this phase, and of opposite sign to B-NoFb (see Table \ref{table:EmagGrowth}).

Finally, in comparison with isolated simulations of MW galaxies \citep{Rieder16}, where there is no accretion phase, cosmological simulations seem to display a lower growth-rate $\Gamma_{\varepsilon_\text{mag}}$. This is especially interesting in the case of B-RdTh, as this run features a similar subgrid feedback as one of the isolated galaxy runs by \citet{Rieder16}. Small disparities could arise from a combination of a slightly diminished feedback specific energy injection and the subdominant cosmic accretion of poorly magnetised gas, which will act to somewhat reduce our $\Gamma_{\varepsilon_\text{mag}}$. However, the most likely cause of significant discrepancy is that their fixed gravitational potential, where no fresh gas is accreted after the initial collapse phase, facilitates the escape of outflows from the galaxy far into the halo, thus promoting the establishment of a galactic fountain able to stretch (and therefore amplify) the field more efficiently than in our cosmological set-up.

\section{Discussion}
\label{s:Robust}

The main result of this paper is the identification of three distinct phases of magnetic energy amplification during a galaxy lifetime. In this section, we discuss the robustness of this result. While there are physical arguments that support the idea that such an amplification behaviour in three periods might be universal, the employed numerical prescriptions affect the duration and amplification occurring during the phases in various manners.

Firstly, a collapse driven phase of amplification is identified separately from the later evolutionary phases of the galaxy. It is constrained to occur between the turn-around time of the collapsing perturbation and the moment when the forming galaxy has generated its first stars. We measure these two moments as $z_\text{ta} \sim 22$ and $z_\text{coll} \sim 13$. The latter redshift strongly correlates with a change of behaviour in the magnetic energy amplification, which during the collapse phase, is mostly the result of frozen magnetic field  lines being dragged by the collapsing gas. Motions are dominated by gravity, and magnetisation levels are comparable across all three simulations, with the stellar feedback subgrid implementation
having a noticeable, but subdominant effect. While the existence of such a phase is difficult to fault, this does not mean that the amplification is unaffected by numerical considerations. In fact, two aspects have considerable impact on the amplification found during the collapse phase. One is numerical resolution, which is discussed in more detail in appendix \ref{ap:resolution}. It mainly affects the amount of turbulent amplification measured above the expected level for isotropic collapse. The second one is the orientation and structure (i.e. spatial variation) of the initial seed. This aspect is not explored here, and will be further analysed in Martin-Alvarez et al. (in preparation), but we note that such variation of the initial seed field could alter the amplification factor of the specific magnetic energy by about a dex. 

After this initial collapse, galaxies rapidly develop a multiphase ISM and are mostly fed by cold filamentary gas flows \citep[see][]{Tillson15}. This second phase is identified as the accretion phase of amplification. Turbulent motions in the multiple phase ISM are deemed responsible for the growth of the magnetic field. This implies the presence of turbulent dynamo amplification, for which $E_\text{mag}$ follows an exponential growth. During this epoch, we only measure a weak rotational support of the gas and any differential rotation amplification should therefore be negligible. Per contra, a turbulent dynamo fed by resolved vortical motions should exponentially amplify the magnetic energy. Filamentary accretion flows are expected to naturally drive solenoidal turbulence, but whether they can transfer a significant amount of turbulent kinetic energy to the entire ISM is still a matter of debate. However, we find that whenever implemented, stellar feedback, while not preventing amplification of magnetic energy to occur, appears to be detrimental to it. As a consequence, the turbulent amplification found during this phase is primarily attributed to accretion flow driven gravitational instabilities. As amplification is linked to turbulence, a higher numerical resolution will yield considerably higher growth rates by pushing down the numerical dissipation scale and resolving a larger span of the turbulent cascade. The typical growth rates of specific magnetic energy that we measure during this phase are on the order of $\Gamma_{\varepsilon_\text{mag}} \sim 2 \text{ Gyrs}^{-1}$ (see Table \ref{table:EmagGrowth}), and are relatively independent of the feedback prescription employed. 

The beginning of the accretion phase coincides with the end of the collapse phase $z_\text{coll} \sim 13$, itself determined by cosmological (and star formation) arguments, but turbulent amplification inside the proto-galaxy is likely to have already started before this redshift. 
As time progresses, the mass growth of the galaxy and the decrease in gas accretion rate by the galaxy steadily reduces the relevance of cold accretion flows, eventually leading to the termination of accretion-driven magnetic energy amplification. In the absence of other turbulence driving mechanisms (B-NoFb run), any amplification is likely confined to the outer regions of the galaxy and an overall decay of the specific magnetic energy is observed. This decay is, by-and-large, the result of numerical resistivity. However, galaxies are expected to host energetic phenomena, such as stellar feedback, stirring up the ISM. The diminishing importance of accretion-driven turbulence thus leads to a smooth transition to a feedback-dominated regime. In our simulations, this latter is observed to take place after the galaxy mass becomes star dominated and shortly after the formation of a thin gas disk. In fact, one 
could even argue for the presence of an intermediate transition phase once the galaxies become star dominated (especially in the absence of feedback), when the stability of the gas disk is enhanced by the presence of a dominant stellar component, and the growth rate is slightly reduced. This should be interpreted with caution though, due to the complication of having to disentangle merger/interaction effects with external objects. The transition from accretion to feedback driven amplification is estimated to occur at $t_\text{acc/fb} \sim 1.7$ Gyrs (or equivalently, $z_\text{acc/fb} \sim 4$).

The final phase of amplification, which we dub the feedback phase of amplification is characterised by lower amplification rates. Tables are now turned and stellar feedback is the main mechanism responsible for the amplification. As a result, the B-NoFb run, which displayed the highest growth rate now exhibits a magnetic energy decay while the B-RdTh run shifts from the lowest growth rate to the highest. As for the accretion dominated phase, the turbulent nature of the amplification implies that higher numerical resolution will better capture the flow turbulence and lead to higher growth rates. A large spread of growth rates are found for this phase, which depend sensitively on feedback implementation $-2.5 \text{ Gyrs}^{-1} < \Gamma_{\varepsilon_\text{mag}} < 1.5 \text{ Gyrs}^{-1}$.

\begin{figure}%
\includegraphics[width=\columnwidth]{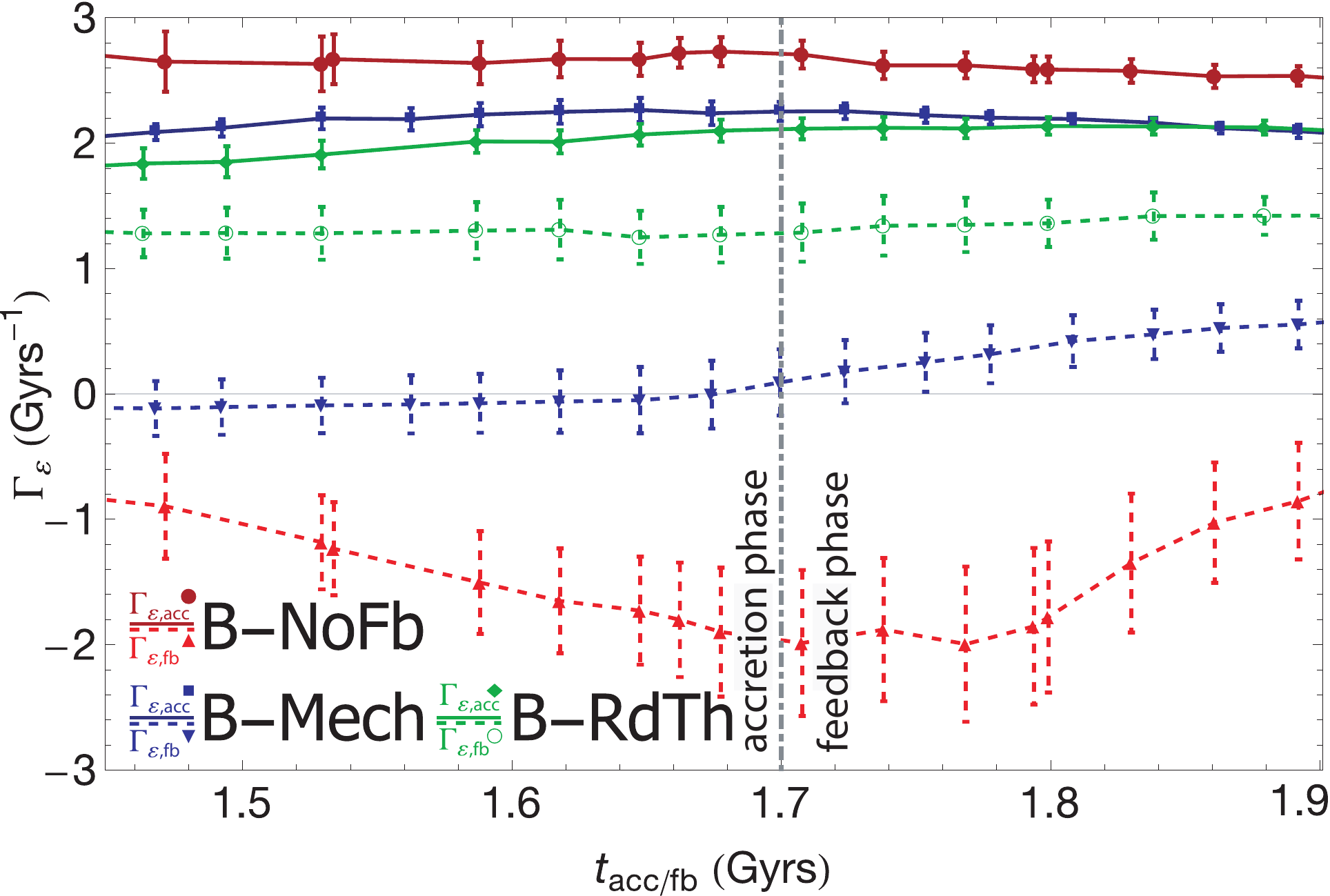}
\caption{Measured exponential growth parameter $\Gamma_{\varepsilon_\text{mag}} $ versus time of phase transition $t_\text{acc/fb}$ for the three runs B-NoFb (red), B-Mech (blue) and B-RdTh (green). Solid coloured lines indicate accretion phase values $\Gamma_{\varepsilon_\text{mag}, \text{acc}}$, while dashed lines represent feedback phase values $\Gamma_{\varepsilon_\text{mag}, \text{fb}}$. Note that  the distinction between phases is robust regardless of data binning or specific choice for the time at which the phase transition takes place.}%
\label{Robustness}%
\end{figure}

In fine, the distinction between accretion and feedback dominated phase is thus underpinned by the robustness of our $\Gamma_{\varepsilon_\text{mag}} $ measurements. In Table \ref{table:EmagGrowth}, best fits of this parameter to magnetic growth time scales are shown. These are calculated employing 7 bins to fit data spanning either the entire duration of the accretion phase or that of the feedback phase, with an exponential function $A \exp{\left[\Gamma_{\varepsilon_\text{mag}} t\right]}$. For each phase, data is smoothed using a 0.2 Gyr window function to minimise the contribution of sharp features and better recover the smooth background behaviour of the curve. As the end of the collapse phase is considered to be clearly defined, $A$ is fixed to its value $\varepsilon_\text{mag,C}$ for the accretion phase. It is left as a free parameter for the feedback phase. To probe the robustness of the transition between the two phases, in Figure \ref{Robustness} we display the different values of $\Gamma_{\varepsilon_\text{mag}}$ we obtain when varying $t_\text{acc/fb}$. We find that modifying the binning of the data or the size of the window function has a negligible impact on the fitted values of $\Gamma_{\varepsilon_\text{mag}}$, but that they are more sensitive to the choice of the epoch at which the phase transition takes place, $t_\text{acc/fb}$. This is not very surprising, as varying $t_\text{acc/fb}$ results in the inclusion of a larger fraction of the 'wrong' phase in the estimate of the value 
of $\Gamma_{\varepsilon_\text{mag}}$. However, as shown by Fig. \ref{Robustness}, the amount of uncertainty is quite limited:  values of $\Gamma_{\varepsilon_\text{mag}}$ remains fairly constant over a reasonable range of $t_\text{acc/fb}$ for all phases and all runs. Even for the phase 
where the estimate $\Gamma_{\varepsilon_\text{mag}}$ varies the most (feedback phase of the B-NoFb run, dashed red line of Fig. \ref{Robustness}), we measure a clear decay of the magnetic energy, in stark contrast with a clear amplification in the matching accretion phase (solid red line on the same plot).

The amplification of the specific magnetic energy reached in our galaxies by $z = 2$ becomes all the more remarkable when considering that these latter mainly grow via low-magnetised gaseous inflows, and lose magnetic energy via outflows. Contrary to isolated simulations of galaxies \citep[e.g.][]{Dubois08,Rieder16}, stellar feedback need not necessarily be the dominant component which drives magnetic amplification in cosmological simulations. 
However, once accretion becomes unimportant, these idealised simulations resemble more closely to the cosmic scenario, and stellar feedback probably plays an important role in preventing the decay of the magnetic energy or generating further amplification. In any case, whatever the driving process, numerical resolution is the decisive factor to obtain the correct rate of  amplification \citep{Rieder17b}. Extrapolating our numerical results to physical ISM magnetic resistivities (and assuming non-ideal MHD effects on small scales such as ambipolar diffusion do not play a major role), would lead to saturation of the magnetic energy in timescales shorter than a Gyr, i.e. within the accretion phase.

\section{Conclusion}
\label{s:Conclusion}
In this work, we performed high resolution cosmological zoom-in simulations of a Milky-Way like galaxy \citep[extending the {\sc nut} suite to MHD, see][for detail about the original simulations]{Powell11} with the  {\sc ramses-mhd} code \citep{Teyssier02,Fromang06, Teyssier06}, which uses a constrained transport scheme to solve the magnetic field induction equation and therefore guarantees an as low a divergence of the magnetic field as numerically possible. 

We studied the evolution of the magnetic energy within this galaxy starting from an extremely weak primordial seed field and using three simulations dubbed B-NoFb, B-Mech and B-RdTh where the only difference between simulations is the stellar feedback sub-grid model implemented.
We find that:
\begin{itemize}
\item the evolution of the magnetic energy can be decomposed as a smooth exponential amplification/decay component on top of which sharp spikes, corresponding to merger events, superimpose. 

\item the amplification of the magnetic energy occurs in three main phases: an initial collapse phase, an accretion-driven phase, and a stellar feedback-driven phase. The two latter amplification phases occur because of the presence of a turbulent dynamo.

\item during the collapse phase, magnetic energy amplification closely follows an isotropic adiabatic approximation, but is already accompanied by turbulent motions which drive a larger growth of the magnetic field.  
\item the accretion-driven phase accounts for a significant fraction of the amplification, and does not require the presence of stellar feedback to sustain the turbulent dynamo. As a matter of fact,  during this phase, feedback is found to be {\em detrimental} to amplification, as the B-NoFb run reaches the highest amplification level. 

\item the contribution of mergers to the overall amplification is negligible. Mergers trigger temporary enhancement of the field, primarily by stretching magnetic field lines. In the absence of stellar feedback, these lines are not disrupted and generate sharp spikes of amplification, but once merging events conclude, the field lines fall back onto the galaxy and most of the corresponding magnetic energy is lost through numerical reconnection. 

\item as the galaxies grow in size and cosmic accretion dwindles, their disks stabilise and, in the absence of stellar feedback, the level of turbulence decreases and the specific magnetic energy decays. This typically occurs around $z \sim 4$.

\item in the runs where stellar feedback is present, it is able to drive a significant turbulent dynamo {\em after} the accretion phase, which further amplifies the field. The most efficient feedback implementation (B-RdTh run), which was the most detrimental during the accretion phase, sustains the highest growth of specific magnetic energy. 
\end{itemize}

Although we have performed quantitative measurements of the specific magnetic energy growth rates, $\Gamma_{\varepsilon,\text{acc}}$, 
the numerical values reported in the present work should only be considered as lower limits. This is because the numerical resolution, 
which determines the magnetic resistivity (as well as the shear and bulk viscosities) of the gas flow is necessarily limited. 
We therefore expect that, as the resolution is increased, the value of this latter will drop and the growth rate of magnetic energy will rise accordingly. 
We provided a first attempt at quantify this resolution effect in appendix \ref{ap:resolution}, which we believe is somewhat encouraging. 
However, further work is needed to properly establish the level of convergence we have reached, given the complex interplay between the numerics, multiphase ISM and implemented subgrid physics. 

This difficulty notwithstanding, we can compare our results to other, similar numerical approaches.
By the end of the collapse phase, the specific magnetic energy of our galaxies is found to lie a factor $\sim 5 \text{-} 10$ above 
the level predicted by isolated galaxy simulations or analytical estimates. 
Using a similar setup to our B-RdTh run, albeit to simulate a smaller dwarf galaxy, \cite{Rieder17b} measure, in the entire zoomed region, 
a growth rate for the magnetic energy density $\Gamma_{\epsilon,\text{acc}}$ which is an order of magnitude larger than that in our galactic region during the accretion phase.On the other hand, the magnetic energy growth during the feedback phase in our simulation only is slightly weaker than in isolated spirals simulations by \cite{Rieder16}. 

Finally, given the level of resolution we can currently achieve, a much higher amplitude of the initial cosmic seed field, or an injection by stellar sources within the galaxy \citep[as in e.g.][]{Butsky17} are needed is needed to reach saturation of specific magnetic energy through turbulent galactic dynamo amplification. We plan to address the potential impact of the back-reaction induced by such a saturated field on the three-phase amplification of cosmic magnetic fields in future work.

\section*{Acknowledgements}
The authors kindly thank the referee for insightful comments and suggestions that contributed to improve the quality of this manuscript. This work was supported by the Oxford Centre for Astrophysical Surveys which is funded through generous support from the Hintze Family Charitable Foundation. SMA thanks M. Rieder for useful discussions and hospitality during his visit to UZH. This work is part of the Horizon-UK project, which used the DiRAC Complexity system, operated by the University of Leicester IT Services, which forms part of the STFC DiRAC HPC Facility (\href{www.dirac.ac.uk}{www.dirac.ac.uk}). This equipment is funded by BIS National E-Infrastructure capital grant ST/K000373/1 and STFC DiRAC Operations grant ST/K0003259/1. DiRAC is part of the National E-Infrastructure. The authors would like to acknowledge the use of the University of Oxford Advanced Research Computing (ARC) facility in carrying out this work. \href{http://dx.doi.org/10.5281/zenodo.22558}{http://dx.doi.org/10.5281/zenodo.22558}. The authors also thank Taysun Kimm for sharing his mechanical feedback prescription and acknowledge the usage of the FFTW library: \href{http://www.fftw.org/}{http://www.fftw.org/}.




\bibliographystyle{mnras}
\bibliography{./references}



\appendix

\section{Resolution tests}
\label{ap:resolution}
As mentioned in the main body of the text, one of the factors that is expected to have a significant impact on the amplification of magnetic energy throughout the entire simulation is numerical resolution. The higher the resolution, the smaller the scale that can be captured by the simulation, and the larger
the turbulence cascade that can be captured. Smaller scale turbulence has smaller e-folding times and therefore yields faster amplification. At the same time, the numerical viscosity, which also scales with resolution, is reduced and turbulence is thus dissipated at larger wavenumbers, increasing the inertial range. Finally, the numerical magnetic resistivity decreases as well, and the decay of the magnetic field is thus less pronounced.

\begin{figure}%
\includegraphics[width=\columnwidth]{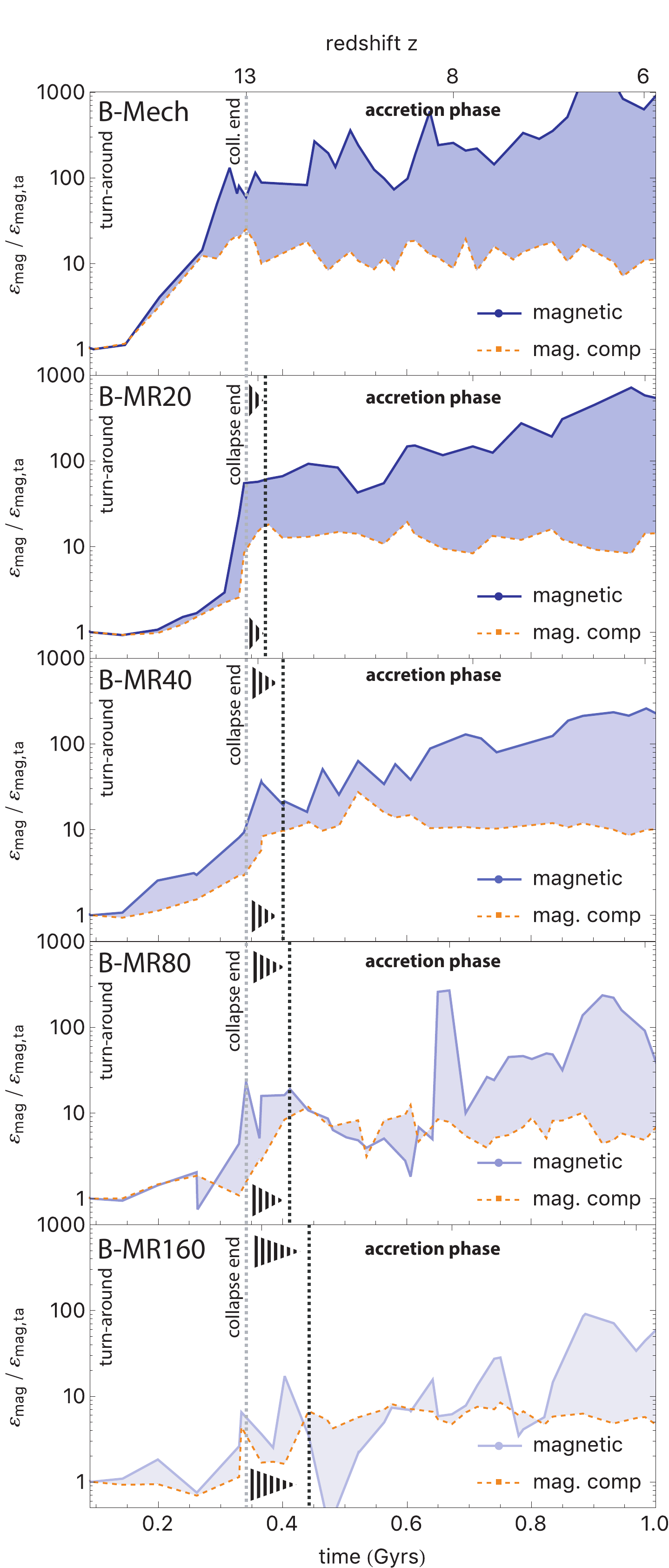}
\caption{Specific magnetic energy growth during the collapse phase and the early stages of the accretion phase. The five runs plotted 
are identical B-Mech simulations with decreasing spatial resolution $\Delta x_\text{m} = $ 10, 20, 40, 80 and 160 pc from top to bottom respectively. The approximate moment of collapse is delayed as AMR levels are removed. The amplification rate $\Gamma_{\varepsilon_\text{mag}}$ is reduced as the turbulence is less and less resolved (Table \ref{table:ResolutionGrowth}).}
\label{Resolution}%
\end{figure}

As a simple way to quantify the importance of this resolution effect, five B-Mech like runs (Table \ref{table:setups}) that differ exclusively in their maximum spatial resolution (B-Mech, B-MR20, B-MR40, B-BMR80, and BMR160 with $\Delta x_\text{m} = $ 10, 20, 40, 80, and 160 pc respectively) are compared in Fig. \ref{Resolution}. This figure is an analogue of Fig.\ref{EarlyEnergy}, but for different spatial resolutions instead of feedback prescriptions. The adiabatic isotropic amplification estimate of the specific magnetic energy for the four runs with the highest resolution reaches a similar value ($\sim 10\varepsilon_\text{ta}$), but appears slightly inferior for the lowest resolution ($\sim 7\varepsilon_\text{ta}$). The collapse phase is completed around $\sim 0.35$ Gyrs for the B-Mech run and seems to be delayed no more than $0.05$ Gyrs for the four highest resolutions, while it is delayed $\sim 0.1$ Gyrs for the lowest resolution (the gravitational force softening scales with spatial resolution). As expected, both the initial collapse amplification and the later turbulent growth are augmented as resolution increases, although there seems to be a reasonable level of convergence between B-Mech and B-MR20. We expect this to be the result of only a fraction of the galaxy being resolved at $10$ pc in B-Mech (see Fig. \ref{GalRes}). The growth parameters measured for this early accretion phase, $\Gamma_{\varepsilon_\text{mag},\text{early}}$ are shown in Table \ref{table:ResolutionGrowth}. As expected, increasing spatial resolution leads to more rapid amplification rates \citep{Rieder16,Pakmor17}.

\begin{figure}%
\includegraphics[width=\columnwidth]{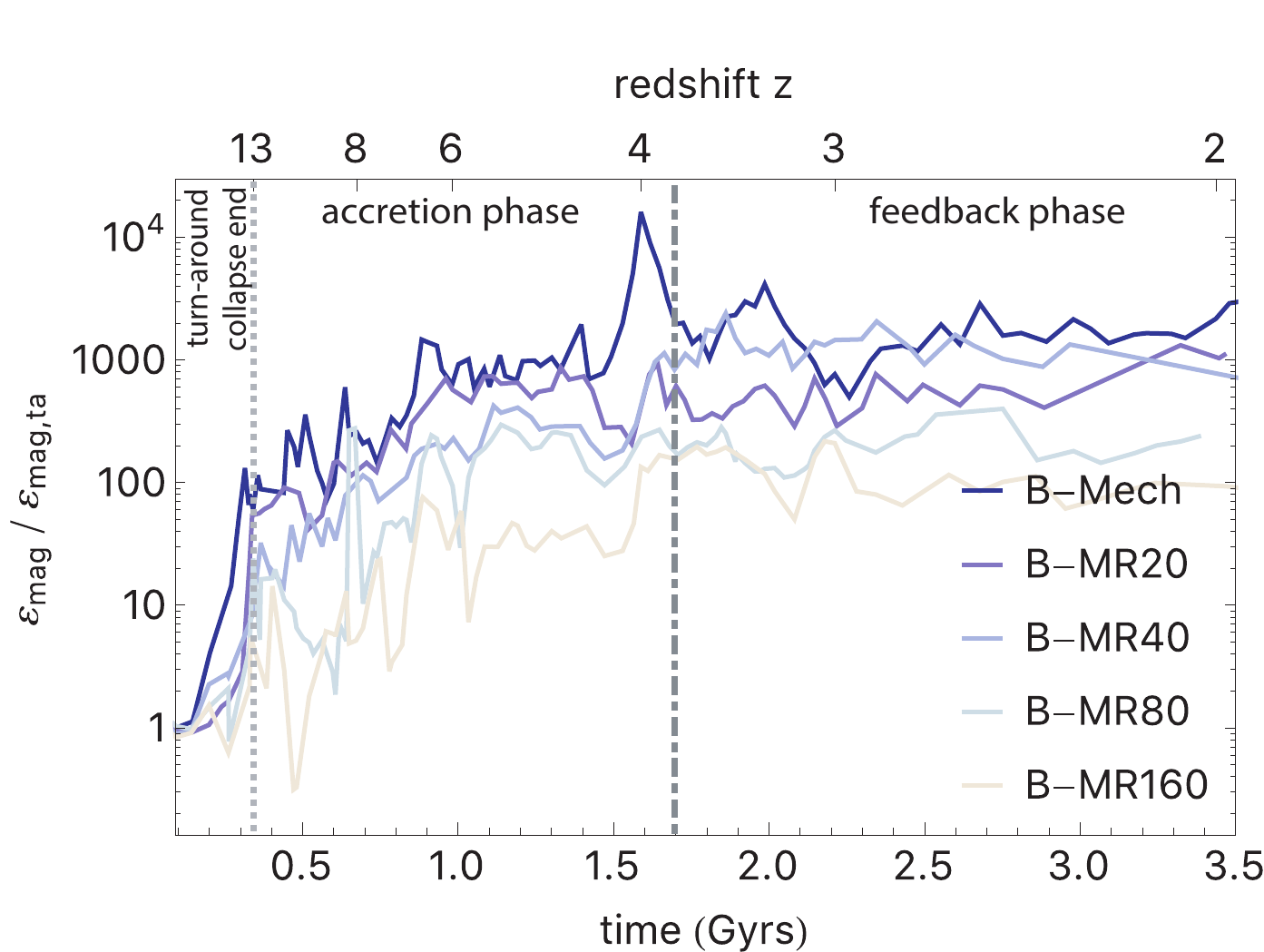}
\caption{Specific magnetic energy growth in the galactic region. The lines represent runs with decreasing resolution from dark blue (B-Mech) to cr\`eme (B-MR160). Regardless of the employed resolution, the three-phase evolution is recovered. B-MR160 appears to be a limiting case, displaying almost no growth. The displayed phases and divisions are the same as shown in Fig. \ref{SpecificEmag}.}
\label{ResolutionPhases}%
\end{figure}

To illustrate the robustness of the accretion and feedback phases with resolution, Fig. \ref{ResolutionPhases} presents the evolution of the specific magnetic energy at different spatial resolutions for the whole duration of the B-Mech simulation. The existence of the accretion and feedback phases appears to be quite insensitive to the numerical resolution. Once again, growth rates $\Gamma_{\varepsilon_\text{mag}}$ measured for the 
different phases are shown in Table \ref{table:ResolutionGrowth}. These values are calculated in the same way as those reported in Table \ref{table:EmagGrowth}. Note that during the feedback phase only B-Mech and B-MR20 possess enough resolution to capture turbulent dynamo growth. 
In B-MR40 and B-MR160 the magnetic field decays. Similarly, the growth factor for B-MR80 is small and compatible with null.

\begin{table}
\centering
\caption{Specific magnetic energy exponential fits for $\Gamma_{\varepsilon_\text{mag}}$ in the five resolution runs. Values are for the period displayed in Fig. \ref{Resolution} (early), and the accretion (acc) and feedback (fb) phases. All $\Gamma_{\varepsilon_\text{mag}}$ timescales are measured in Gyrs$^{-1}$. Calculations of the growth rates are as for Table \ref{table:EmagGrowth}. Last column displays a bulk estimate of the magnetic Reynolds number $Rm$ in the galaxy (see text). The estimated $Rm$ is above or similar to the critical magnetic Reynolds number required for turbulent amplification $Rm_\text{crit} \simeq 30$ in all runs but B-MR160.}
\label{table:ResolutionGrowth}
\begin{tabular}{l | r r r r}
\hline

Run & $\Gamma_{\varepsilon_\text{mag},\text{early}}$ & $\Gamma_{\varepsilon_\text{mag},\text{acc}}$ & $\Gamma_{\varepsilon_\text{mag},\text{fb}}$ & $Rm$ \\
\hline
B-Mech & $2.3 \pm 0.2$ & $2.3 \pm 0.1$ & $0.3 \pm 0.1$ & 160\\
B-MR20 & $2.4 \pm 0.1$ & $1.9 \pm 0.1$ & $0.6 \pm 0.3$ & 120\\
B-MR40 & $2.0 \pm 0.1$ & $1.9 \pm 0.1$ & $-0.2 \pm 0.1$ & 60\\ 
B-MR80 & $1.0 \pm 0.2$ & $1.2 \pm 0.1$ & $0.1 \pm 0.2$ & 30\\
B-MR160 & $0.7 \pm 0.1$ & $0.9 \pm 0.1$ & $-0.3 \pm 0.1$ & 15\\
\hline
\end{tabular}
\end{table}

Turbulent amplification also depends on the value of magnetic Reynolds numbers. \citet{Haugen04,Brandenburg05} indicate that a magnetic Reynolds number $Rm$ above a critical value $Rm_\text{crit} \sim 30-35$ is required in order to trigger the a turbulent dynamo given $Pm = 1$. The critical Reynolds number is found to be lower for the range of $Pm$ values expected in our simulations \citep[see eq. 10 and Fig. 2][$Rm_\text{crit} (Pm) \sim 35 Pm^{-1/2}$]{Haugen04}.

If we define a local magnetic Reynolds number as a function of the local diffusivity $\eta$, and typical length $L_\text{sys}$, and velocity $V_\text{sys}$ of the studied system

\begin{equation}
Rm = \frac{L_\text{sys} V_\text{sys}}{\eta},
\label{eq:magReynolds}
\end{equation}

we can compute a bulk estimate of the magnetic Reynolds number in the ISM of the galaxy. In order to do that, we fix the scale length of the system to be equal to the entire thickness of the disk $L_\text{sys} = 2 h_s \sim 1.2 \text{kpc}$. Note that $h_s$ is in the range $0.6\, \text{kpc} - 0.4\, \text{kpc}$ between redshifts $z = 8$ and $z = 2$, and that this scale is expected to somewhat increase with decreasing resolution.

\citet{Teyssier06} argue that the amount of local numerical diffusivity introduced by the Godunov solver in {\sc ramses} is on the order of $\eta \sim 0.5 | u | \Delta x$, with $|u|$ the (1D) flow velocity, and $\Delta x$ the size of a resolution element. Assuming that $|u| \sim V_\text{sys}$, we can compute an approximate magnetic Reynolds number $Rm \sim Rm (\Delta x) = 2 L_\text{sys} / \Delta x$. Under these assumptions our magnetic Reynolds numbers (presented in Table \ref{table:ResolutionGrowth}) fulfil $Rm \gtrsim Rm_\text{crit}$ for all runs but B-MR160. Accordingly, we expect some degree of turbulent amplification for all the different resolution runs but B-MR160, and possibly B-MR80. This seems in good agreement with the growth parameters presented in Table \ref{table:ResolutionGrowth}.

Growth rates likewise depend on turbulence properties such as Mach number $\mathcal{M}$, or Helmholtz mode fraction. For significantly higher magnetic Reynolds numbers, \citet{Federrath11b} find purely solenoidal turbulence forcing to produce higher growth rates than compressional forcing. Nonetheless, they find amplification for both types of forcing, specially in the case of supersonic turbulence. We thus expect a similar behaviour for our simulations, with $\left<\mathcal{M}\right> \sim 10^{1.5}$ and a typical fraction of solenoidal turbulence $\sim 50 - 70\%$ in the galactic region. We also expect that the scaling of their supersonic growth rates $\propto \mathcal{M}^{1/3}$ applies to our simulations.

This confirms the three-phase character of the turbulent dynamo amplification during its purely kinematic regime. Further increase of resolution is expected to lead to a higher level and more rapid amplification.

\section{Fast Fourier Transforms of a galaxy}
\label{ap:diskSpec}

In this appendix the characteristics of the spectrum of a disk galaxy embedded within a cubic box are addressed, as this is used extensively to interpret results in the main body of the text and disentangle physical properties from numerical artefacts in a given spectrum. We caution that our approach does not fully differentiate features caused in the spectral analysis by the gas density structure. A more localised analysis based, e.g. on wavelets, would be required to fully characterise which of the spectral features are exclusively generated by turbulence. The aim of this section is to provide a better understanding of the overall impact of the morphology of the galaxy. More specifically, one assumption commonly made in the calculation of a FFT is that of domain periodicity. While it is not adequate for the case of an individual galaxy, an accurate spectrum can still be calculated if precautions are taken. Similarly, it is important to discriminate the galaxy from its environment in a cosmological simulation and to recognise features associated with its overall shape (disk scale height and length) instead of the dynamics of the gas flow within it.

First, we have to discriminate the studied object from its environment. This occurs by definition in isolated simulations of galaxies, where all the effects only are the result of the presence of a galaxy. Fortunately, physical quantities (i.e. kinetic and magnetic energies) have values in the galaxy several orders of magnitude above the ones found in its environment. This naturally acts as an effective 0-padding: the Fourier modes outside the galaxy are negligible with respect to the amplitude obtained for the ones within the galaxy, as long as the wavelength remains smaller than the size of the region of interest. In physical terms, the energy $E(k)$ contained in a given scale $k$ smaller than the scale of the galaxy $k > k_\text{gal}$ is found to be negligible in the environment compared with the energy corresponding to the same mode within the galaxy. Accordingly, the studied scale $k$ has to be smaller than the largest scale corresponding to the galaxy. As the size of the studied galaxy roughly scales with the size of the box (with a side length $l = 0.2\, r_\text{halo}$), and the radial scale of the galaxy is typically $\sim 0.1\, r_\text{halo}$, this corresponds to $k_\text{gal} \sim 2\, k_\text{min}$, where $k_\text{min}$ stands for the minimum spectrum scale represented. Smaller $k$s  should 
therefore be discarded from the analysis even though we still display them on the figures. 

In addition, wavenumber scales $k$ on the order of the overlapping of the galaxy with its periodic occurrences should be neglected. Correlations can be found between modes that bridge the shortest distance between the studied object and its periodic analogue. For the aforementioned size of the box and galaxy, the maximum scales $k$ that can be affected by periodicity are again $k_p \sim 2\, k_\text{min}$.

\begin{figure}%
\includegraphics[width=\columnwidth]{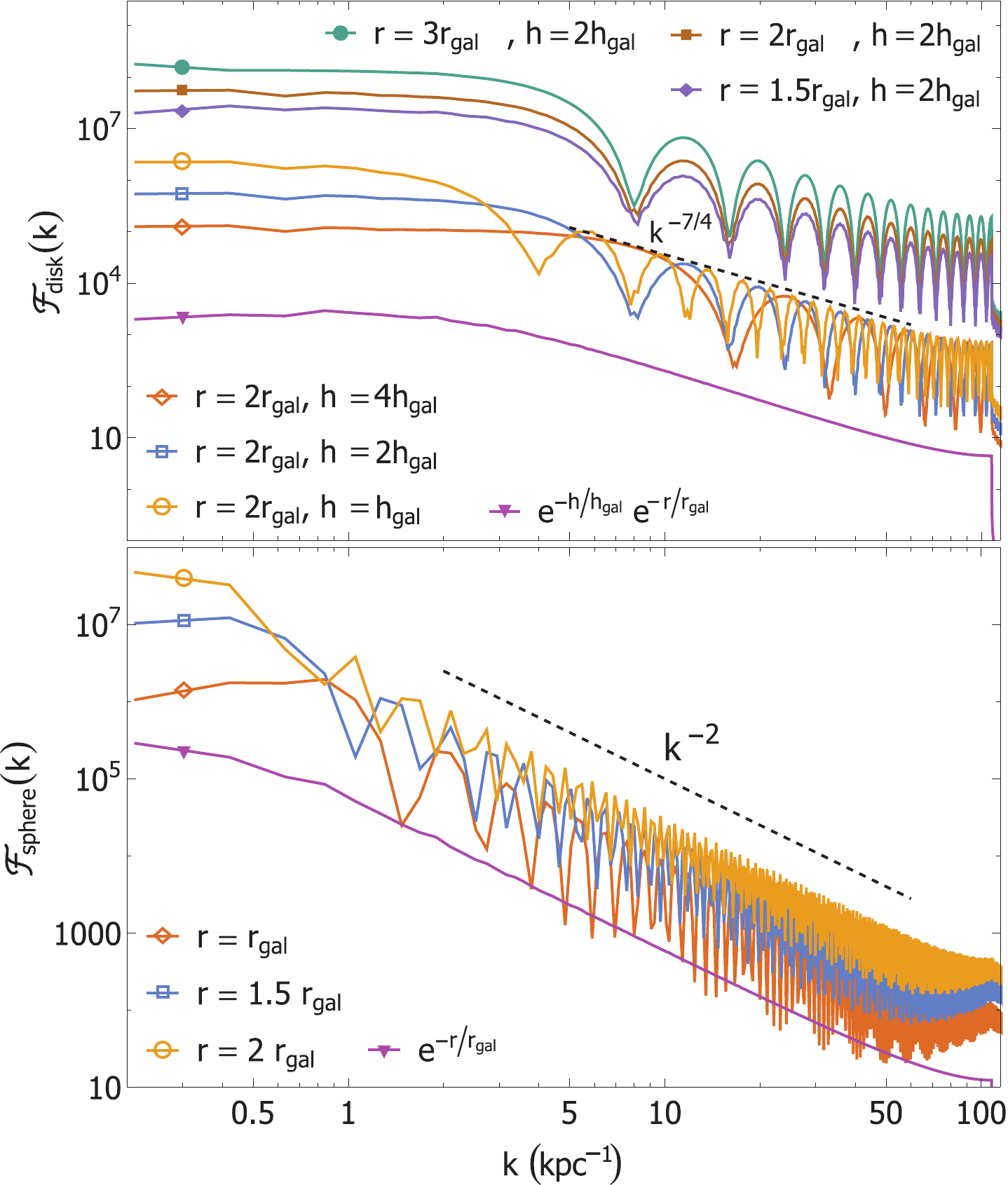}
\caption{Spectra for fiducial shapes of a galaxy. The impact of the shape parameters is assessed by plotting different thicknesses and radii for a disk (top) and a sphere (bottom). The three top lines for a disk have been shifted a factor $\times 1000$ for the sake of clarity. Purple lines display the spectrum corresponding to an exponential density profile rather than a uniform one, as expected for a more realistic galaxy. The dashed black lines display the exponent of the power-law decay: $\mathcal{F}_\text{disk} (k) \propto k^{-7/4}$ for a disk and $\mathcal{F}_\text{sphere} (k) \propto k^{-2}$ for a sphere.}
\label{ShapeSpec}%
\end{figure}

Due to the small volume of the box covered by the galaxy, the next consideration is the actual shape of the galaxy. It can be simplified to either a sphere during early stages of formation and a thick disk otherwise. For the case of a disk, a uniform density disk of half-thickness $h = \alpha h_\text{gal}$ and polar radius $r_\text{polar} = \beta r_\text{gal}$ is studied, where $h_\text{gal} \sim 200$ pc and $r_\text{gal} \sim 3$ kpc are the half-thickness and radius resulting from fitting the gas density profile of the B-Mech galaxy with an exponential form. Figure \ref{ShapeSpec} displays how the variations of the $\alpha$ and $\beta$ parameters for a disk affect its spectrum. The strong oscillations seen in the spectrum are caused by the disk being implemented as a Heaviside-like window function. The thickness of the disk affects the scale of the first minimum, $k = \frac{\pi}{h}$ (i.e. the beginning of the decline). On the other hand, the radius of the disk affects how sharp the decay to the minimum is and the amount of power contained in the largest scales, as well as the mean value at $k = 0$. As the radius becomes larger, the spectrum gains power on scales $k < 3 k_\text{min}$. Conversely, smaller radii lead to a lowering at $k < 3 k_\text{min}$. As expected, the normalisation of the curves is proportional to the size of the disk, affected by both parameters $\left(\alpha,\,\beta \right)$. Finally, the decay followed by all disk spectra is fitted by $\mathcal{F} (k) \propto k^{-7/4}$ (black dashed line, Figure \ref{ShapeSpec}).

A spherical galaxy shape on the other hand, leads to a more simple spectrum. It contains roughly constant power on the largest scales and starts decaying at the scale of the sphere radius. In Figure \ref{ShapeSpec}, the spectra of a uniform density sphere of radii $r = \alpha r_\text{gal}$ for various $\alpha$ are displayed. The decay of the spectrum for the sphere follows $\mathcal{F} (k) \propto k^{-1.92} \sim k^{-2}$. Changing the radius of the solid sphere exclusively affects the normalisation and once the radius becomes small enough, causes a decay at scales larger than the radius scale due to periodicity effects disappearing.

Finally, to present a more realistic case, both the disk and the sphere are implemented decreasing exponential density profiles
\begin{equation}
v_\text{cyl} (r_\text{polar}, h) = \text{exp}\left(- r_\text{polar} / r_\text{gal} \right) \cdot \text{exp}\left(- h / h_\text{gal} \right),
\label{cylVals}
\end{equation} 
\begin{equation}
v_\text{sp} (r) = \text{exp}\left(- r / r_\text{gal} \right).
\label{spVals}
\end{equation}
Their spectra transform in the same way as their uniform counterparts, but no longer present oscillations.

\begin{figure}%
\includegraphics[width=\columnwidth]{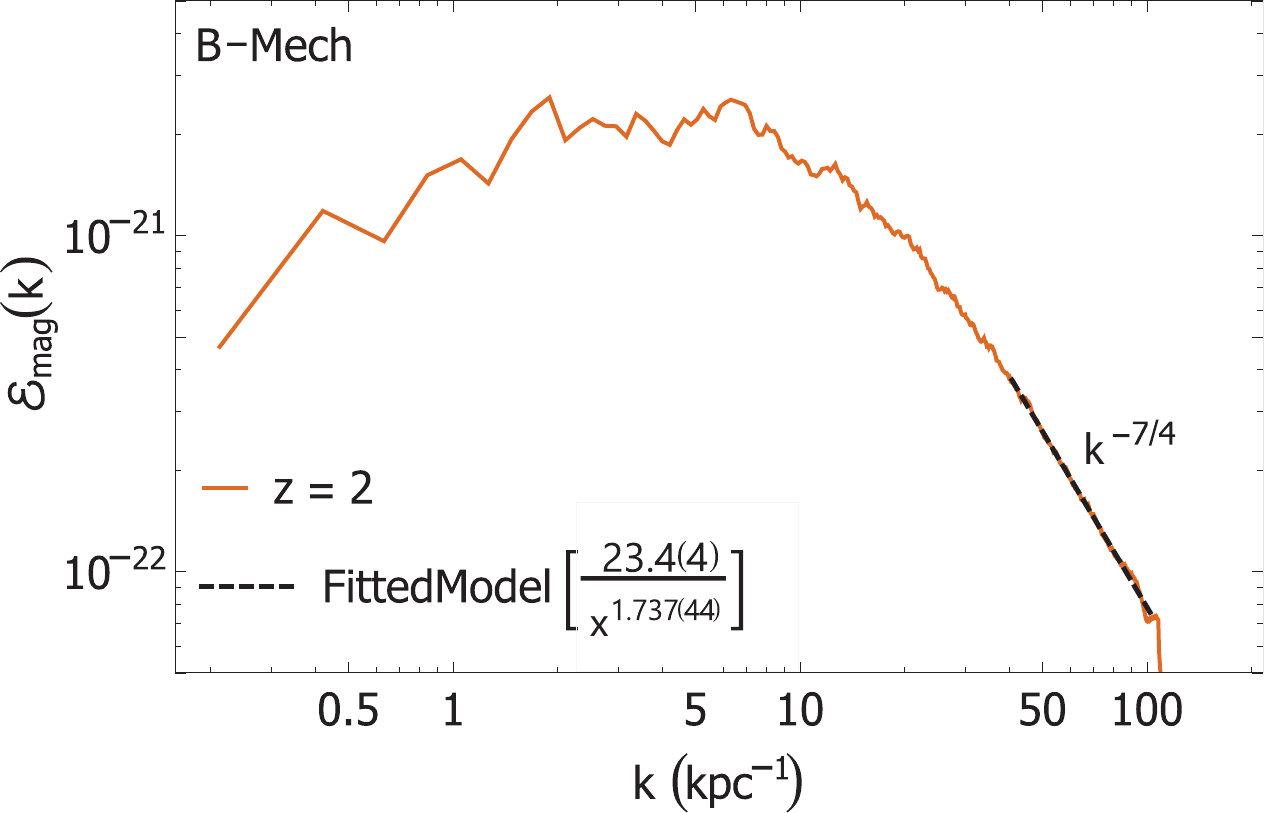}
\caption{B-Mech magnetic energy spectrum at $z = 2$ calculated as indicated in section \ref{ss:AccAmplification}. The black dashed lines represent a power-law fit to the spectrum in the corresponding $k$ region, with the result of the fit shown on the figure ($1\,\sigma$ errors shown in brackets). After $k \sim 10 \text{ kpc}^{-1}$, the spectrum progressively converges to the $\mathcal{E}_\text{mag}\propto k^{-7/4}$ regime expected for an exponential disk (see text for detail). This shape-domination of the spectra is observed for studied spectra whenever magnetic fields are well confined to the galaxy.}
\label{DiskDecay}%
\end{figure}

The measured disk decay $\mathcal{F} (k) \propto k^{-7/4}$ occurs on the smallest scales whenever a disk is present in the simulations. As an illustration, Fig. \ref{DiskDecay} displays the magnetic energy spectrum for B-Mech at $z = 2$. The decay on small scales starts to be dominated by the shape of the galaxy around $k \sim 10 \text{ kpc}^{-1}$ and eventually follows the aforementioned $\mathcal{E}_\text{mag}\propto k^{-7/4}$, as indicated by the dashed black line.

\bsp	
\label{lastpage}
\end{document}